\documentclass[journal]{IEEEtran}
\usepackage{algorithm}
\usepackage{algcompatible}
\usepackage{algpseudocode}
\usepackage[cmex10]{amsmath}
\usepackage{amssymb}

\usepackage[caption=false,font=footnotesize]{subfig}

\usepackage{array}
\usepackage{cite}
\usepackage{enumerate}
\usepackage{float}
\usepackage[pdftex]{graphicx}
\usepackage{makecell}
\usepackage{psfrag}
\usepackage{url}
\newcommand{\CLASSINPUTbaselinestretch}{1.0} % baselinestretch
\newcommand{\CLASSINPUTinnersidemargin}{1in} % inner side margin
\newcommand{\CLASSINPUToutersidemargin}{1in} % outer side margin
\newcommand{\CLASSINPUTtoptextmargin}{1in}   % top text margin
\newcommand{\CLASSINPUTbottomtextmargin}{1in}% bottom text margin

\newtheorem{theorem}{Theorem}
\newtheorem{example}{Example}
\newtheorem{lemma}{Lemma}

\newtheorem{definition}{Definition}
\newtheorem{remark}{Remark}

\def\I{\cal{I}}

\def\C{\mathbb{C}}
\begin{document}

\title{The Feasibility of Interference Alignment for Reverse TDD Systems in MIMO Cellular Networks}
\author{Kiyeon Kim,~\IEEEmembership{Student Member,~IEEE,} Sang-Woon Jeon,~\IEEEmembership{Member,~IEEE,} Janghoon Yang,~\IEEEmembership{Member,~IEEE}\\ and Dong Ku Kim,~\IEEEmembership{Member,~IEEE}
\thanks{K. Kim, J. Yang, and D. K. Kim were supported in part by the IT R\&D program of MOTIE/KEIT [10035389, Research on high speed and low power wireless communication SoC for high resolution video information mining]. S.-W. Jeon was supported in part by the Basic Science Research Program through the National Research Foundation of Korea (NRF) funded by the Ministry of Education, Science and Technology (MEST) [NRF-2013R1A1A1064955].}
\thanks{K. Kim and D. K. Kim are with the School of Electrical and Electronic Engineering,
Yonsei University, Seoul, South Korea (e-mail:
\{dreamofky, dkkim\}@yonsei.ac.kr).}
\thanks{S.-W. Jeon is with the Department of Information and Communication Engineering, Andong National University, Andong, South Korea (e-mail:
swjeon@anu.ac.kr).}
\thanks{J. Yang is with the Department of Newmedia, Korean German Institute of
Technology, Seoul, South Korea (e-mail: jhyang@kgit.ac.kr).}
}
\maketitle
%\markboth{Draft version to be submitted}
%{Shell \MakeLowercase{\textit{et al.}}: MIMO R-TDD cellular networks}

%\IEEEpeerreviewmaketitle


\begin{abstract}
The feasibility conditions of interference alignment (IA) are analyzed for reverse TDD systems, i.e., one cell operates as downlink (DL) but the other cell operates as uplink (UL).
Under general multiple-input and multiple-output (MIMO) antenna configurations, a necessary condition and a sufficient condition for one-shot linear IA are established, i.e., linear IA without symbol or time extension.
In several example networks, optimal sum degrees of freedom (DoF) is characterized by the derived necessary condition and sufficient condition. 
For symmetric DoF within each cell, a sufficient condition is established in a more compact expression, which yields the necessary and sufficient condition for a class of symmetric DoF.
An iterative construction of transmit and received beamforming vectors is further proposed, which provides a specific beamforming design satisfying one-shot IA. Simulation results demonstrate that the proposed IA not only achieve lager DoF but  also significantly improve the sum rate in the practical signal-to-noise ratio (SNR) regime.
\end{abstract}
%%%%%%%%%%%% Introduction %%%%%%%%%%%%
\section{Introduction}
\IEEEPARstart{A}{s} various types of wireless devices and networks emerged, wireless communication demands are explosively increasing and becoming diverse for providing an integrated service of voice, data, video, and so on \cite{Astely13, Cisco14:Online}. 
In order to support such demands, standard groups for next-generation mobile communication have actively studied heterogeneous networks (HetNets) consisting of different types communication systems and devices with different capabilities \cite{3GPPRel12:2,3GPPRel11:2,Kountouris13,Andrews13,Hosseini13,Adhikary14,Ghosh12}. 
In particular, dynamic time division duplex (D-TDD) is considered as a promising technology for heterogeneous cellular networks, which dynamically adjusts the portion of uplink (UL) and downlink (DL) frames based on the current communication traffic and environment in a distributed manner by each heterogeneous cell \cite{WunCheol02, Jungnam04, Illsoo09,Hyoungju13,Hongguang14,CHATTERJEE13}.
Based on the same principle, enhanced interference mitigation and traffic adaptation (eIMTA) has been studied in the 3GPP standard group \cite{3GPPRel11}.

Unlike the conventional cellular networks in which UL and DL phases are synchronized over the entire cells, dynamic adaptation of UL and DL in each cell applied in D-TDD and eIMTA essentially requires a \emph{new interference mitigation technique between asynchronous UL and DL cells}. 
Specifically, for a given time slot, UL cells and DL cells may coexist in a network because of asynchronous coordination between cells, which is also referred to as reverse TDD (R-TDD) systems.
For such R-TDD systems, due to limited capabilities of terminals or users, interference mitigation from the users in UL cells to the users in DL cells is quite challenging, but crucially important for boosting spectral efficiency.
\IEEEpubidadjcol
In the context of D-TDD or R-TDD systems, various interference mitigation techniques have been actively studied in the literature, see \cite{Kountouris13,Andrews13, Hosseini13,Adhikary14,Ghosh12,WunCheol02, Jungnam04, Illsoo09} and the references therein.
For instances, R-TDD techniques has been studied in HetNets to achieve improved area spectral efficiency \cite{Kountouris13} or improved throughputs \cite{Hosseini13, Andrews13,Adhikary14}.
To suppress strong interferences from adjacent users, scheduling policies for D-TDD systems have been studied in \cite{WunCheol02, Jungnam04, Illsoo09}.
The basic principle behind these interference management techniques is to \emph{avoid strong interferences by orthogonalization and treat weak interferences as noise}.

Cadambe and Jafar recently made a remarkable progress showing that the sum degrees of freedom (DoF) of the $K$-user interference channel (IC) is given by $K/2$ \cite{Cadambe08}.
A new interference mitigation paradigm called \emph{interference alignment} (IA) has been proposed to achieve $K/2$ DoF, which align interfering signals from multiple transmitters into the same signal space.
The concept of this signal space alignment has been successfully adapted to various network environments, e.g., see \cite{Cadambe09,Changho11, Changho08,Cadambe09:2,Jeon4:12,Lei12,Tiangao12} and the the references therein.
Different strategies of IA were also developed under the name of ergodic IA \cite{Nazer12,Jeon13,Jeon14:2,Jeon11:2} and real IA \cite{Motahari09:2,Motahari09:1}.
In \cite{Changho08,Changho11}, IA for cellular networks has been studied for both UL and DL scenarios showing that multiple users in each cell are beneficial for improving DoF.
More recently, IA techniques using multiple antennas have been actively studied in order to boost DoF of multiantenna multiuser networks \cite{Jafar07,Tiangao10,Yetis09,Yetis10,  Razaviyayn12,Ruan13,Bresler14,Taejoon11,Duckdong12,Ayoughi13,Vasileios14,Guillaud11,Yanjun12,Tingting13,Tingting13:2,Zhuang11,Jie13,Wonjae11,Joonwoo13,Gokul13,Wonjae12}.
In particular, IA exploiting multiple antennas has been studied in both cellular UL \cite{Taejoon11,Duckdong12,Ayoughi13,Vasileios14,Guillaud11,Yanjun12} and DL \cite{Zhuang11,Jie13,Wonjae11,Joonwoo13,Gokul13,Wonjae12,Tingting13,Tingting13:2}.

In spite of the rapid advances on IA studies for the conventional cellular networks in which the entire cells operate either UL or DL \cite{Taejoon11,Guillaud11,Yanjun12,Duckdong12,Ayoughi13,Vasileios14,Zhuang11,Jie13,Wonjae11,Joonwoo13,Gokul13,Tingting13,Tingting13:2,Wonjae12}, relatively little progress has been made so far on IA for R-TDD systems having the coexistence of UL and DL cells.
One notable work done by Jeon and Suh is to study IA for two-cell R-TDD systems showing that, depending on the antenna configuration, operating one cell as UL and the other cell as DL can enlarge the sum DoF of multiantenna cellular networks than the conventional UL or DL operation \cite{Jeon14arxiv}.
To achieve such DoF gain, asymptotic signal space IA is needed for aligning multiple interfering signals transmitted from the UL users at each of the DL users, which has to apply transmit beamforming over multiple channel instances \cite{Cadambe09,Changho11, Changho08,Cadambe09:2,Lei12,Tiangao12}.

In many practical cellular networks, however, asymptotic signal space IA is hard to implement due to system complexity and feedback overhead. 
Furthermore, it requires large enough channel diversity to improve DoF, which may cause severe delay when the channel coherence time is large.  
In order to overcome such limitations, \emph{single-shot IA using multiple antennas}, i.e., multiantenna beamforming using single time instance, has been considered for the $K$-user IC \cite{Yetis09,Yetis10,Bresler14,Razaviyayn12,Ruan13} and also for cellular networks \cite{Guillaud11,Yanjun12,Tingting13,Tingting13:2}.   
The feasibility conditions of single-shot IA has been first established in \cite{Yetis09,Yetis10} for the $K$-user multiple-input and multiple-output (MIMO) IC having $M$ and $N$ antennas at each transmitter and the receiver respectively.
Followed by \cite{Yetis09,Yetis10}, a more general antenna configuration has been considered in \cite{Razaviyayn12,Ruan13}.
In particular, \cite{Razaviyayn12} established a tighter necessary condition and also the necessary and sufficient condition for a  class of network configurations and  \cite{Ruan13} provided a sufficient condition for general network configurations.
More recently, study on IA feasibility has been extended to MIMO cellular networks for both UL \cite{Guillaud11,Yanjun12} and DL \cite{Tingting13,Tingting13:2}.
IA feasibility conditions have been established in \cite{Guillaud11,Yanjun12} for various UL settings and in \cite{Tingting13,Tingting13:2} for MIMO cellular networks assuming DL. %[KY]
In spite of recent demand on single-shot IA for R-TDD systems along with its significance to heterogeneous cellular networks, the previous work on IA feasibility for MIMO cellular networks inherently assumed cellular UL or DL, i.e., the entire cells operate either UL or DL. 

In this paper, we study \emph{IA feasibility for R-TDD cellular networks having multiple antennas at both base stations (BSs) and users.}
We focus on two-cell environment in which one cell operates as UL but the other cell operates as DL, and restrict to linear IA coding scheme without time or symbol extension as the same reasons assumed in \cite{Yetis09,Yetis10,Bresler14,Razaviyayn12,Ruan13,Yanjun12,Tingting13,Guillaud11}.   
The main contributions of this paper are as follows.

\begin{itemize}[\IEEEsetlabelwidth{Z}]
\item We derive a necessary condition and a sufficient condition on one-shot linear IA for a general MIMO antenna configurations. 
In several example networks, optimal sum DoF is characterized by the established necessary condition and sufficient condition. 
We further demonstrate that the proposed IA improves DoF than that without IA between UL and DL cells.

\item For a symmetric DoF in each cell, we establish a sufficient condition on one-shot linear IA with a more compact expression.
From the sufficient condition derived with a more compact expression, we establish the necessary and sufficient condition for a class of symmetric DoF.

\item We provide an iterative construction method of precoding and postcoding matrices, which provides a specific beamforming design at finite signal-to-noise ratio (SNR). 
Simulation results demonstrate that the proposed IA not only achieve better DoF but also significantly improve the sum rate in the practical SNR regime.
\end{itemize}

The rest of this paper is organized as follows.
In Section \ref{sec:systemmodel}, we introduce the considered MIMO R-TDD cellular network and formally define the IA feasibility problem.
In Section \ref{sec:main_results}, we first state the main results, i.e., a necessary condition and a sufficient condition established in the paper and provide several example networks for better understanding of the main results. % [KY] sufficient -> a sufficient
The detailed proofs of the main results are provided in Section \ref{sec:proof_of_results}.
In Section \ref{sec:IAbuilder}, we propose an iterative construction method for precoding and postcoding matrices and demonstrate by simulation that the proposed construction can improve the sum rate of R-TDD systems in the practical SNR regime. 
We finally conclude in Section \ref{sec:conclusions}. 

%%%%%%%%%%%% Section %%%%%%%%%%%%
\section{Problem Formulation}\label{sec:systemmodel}

In this section, we explain the notation used in the paper and introduce the considered MIMO R-TDD cellular network. We then formally define the feasibility problem for one-shot linear IA.

\subsection{Notation}
Let us introduce the notation used in the paper.
For a matrix $\mathbf{A}$, denote the $i$th row vector and the $(i,j)$th element of $\mathbf{A}$ by ${\mathbf{A}}[i]$ and ${\mathbf{A}} [i,j]$ respectively.
Also,  ${\mathbf{A}}^\dag$, ${\mathbf{A}}^T$, $\|\mathbf{A}\|$,  ${\rm{rank}}( {\mathbf{A}} )$, and ${\rm{det}}( {\mathbf{A}} )$ denote the Hermitian transpose, transpose, Frobenius norm, rank, and determinant of $\mathbf{A}$ respectively.   % [KY] , ${\rm{det}}( {\mathbf{A}} )$ -> , and ${\rm{det}}( {\mathbf{A}} )$
The operator ${\rm{vec}}( {\mathbf{A}} )$ converts $\mathbf{A}$ into the column vector constructed by stacking the column vectors of $\mathbf{A}$, ${\rm{diag}}( {{\mathbf{A}_1}, \cdots ,{\mathbf{A}_n}} )$ denotes the block diagonal matrix whose diagonal blocks are given by ${{\mathbf{A}_1}, \cdots ,{\mathbf{A}_n}}$, and ${\rm{diag}}[ n ]( {\mathbf{A}} )$ denotes the block diagonal matrix whose diagonal blocks are given by ${\mathbf{A}}$ for $n$ times.
The identity matrix of size $n$ is denoted by ${\bf I}_n$ and the all-zero matrix of size $m\times n$ is denoted by ${\bf{0}}_{m\times n}$. 
For a set $\mathcal{A}$, denote its cardinality by $\left| \mathcal{A} \right|$.
The set of complex numbers and the set of natural numbers are denoted by $\C$ and $\mathbb{N}$ respectively. 
Let ${{\rm{mod}}(n,m)}$ denote the modulo operation, i.e., the remainder of $n$ divided by $m$.
The multivariate complex Gaussian distribution with a mean vector $\mathbf{m}$ and a covariance matrix $\mathbf{C}$ is denoted by $\mathcal{CN}(\mathbf{m},\mathbf{C})$.

\subsection{MIMO Reverse Time Division Duplex Cellular Networks}
\begin{figure}			
\begin{center}$
\centering\includegraphics[width=3in]{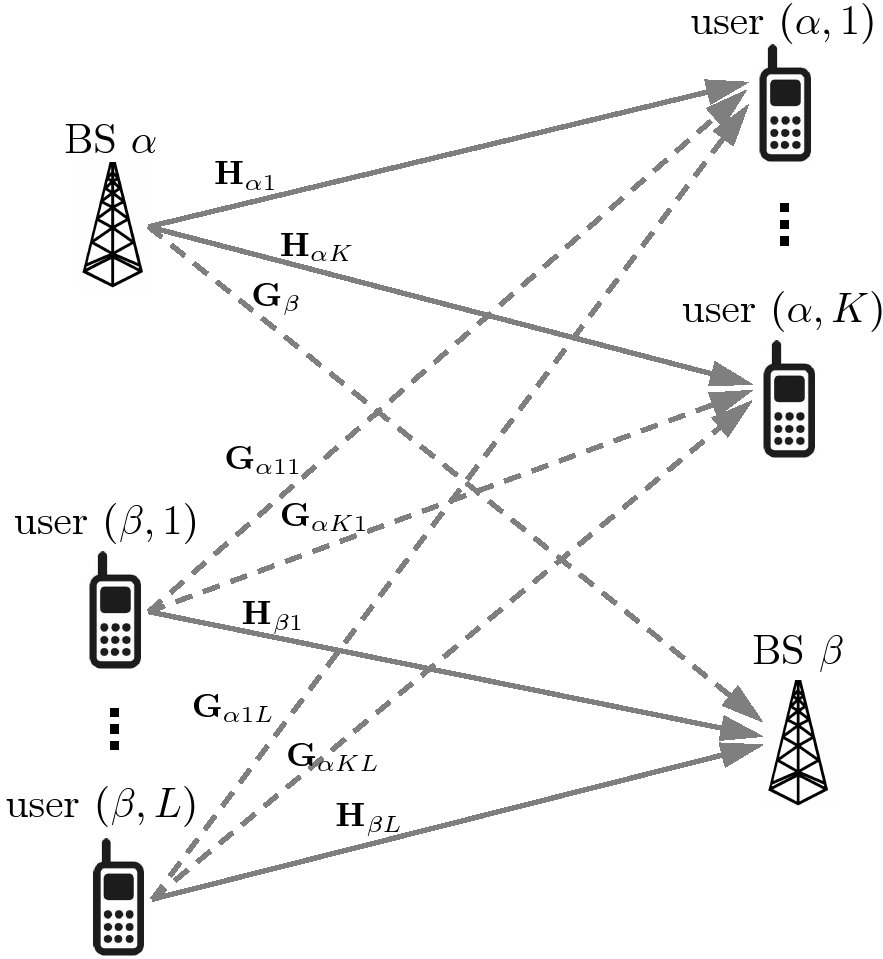}$
\end{center}
\caption{A network model of MIMO R-TDD cellular networks}
\label{fig:system_model}
\end{figure}
Consider a MIMO R-TDD cellular network depicted in Fig. \ref{fig:system_model} in which the first cell (cell $\alpha$) operates as DL, while the second cell (cell $\beta$) operates as UL.
In particular, BS $\alpha$ equipped with $M_{\alpha}$ antennas wishes to transmit independent messages to its $K$ serving users, labeled user $(\alpha,1)$ to user $(\alpha,K)$.
On the other hand, BS $\beta$ equipped with $M_{\beta}$ antennas wishes to receive independent messages from its $L$ serving users, labeled user $(\beta,1)$ to user $(\beta,L)$.
User $(\alpha,k)$ is equipped with $N_{\alpha k}$ antennas and user $(\beta,l)$  is equipped with $N_{\beta l}$ antennas, where $k\in[1:K]$ and $l\in[1:L]$.

The received signal vector of user $(\alpha,k)$ is given by
\begin{equation} \label{eq:output_alpha}
\mathbf{y}_{\alpha k}=\mathbf{H}_{\alpha k}\mathbf{x}_{\alpha}+\sum_{l=1}^L \mathbf{G}_{\alpha kl}\mathbf{x}_{\beta l}+\mathbf{z}_{\alpha k}
\end{equation}
for $k\in[1:K]$ and the received signal vector of BS $\beta$ is given by
\begin{equation} \label{eq:output_beta}
\mathbf{y}_{\beta}=\sum_{l=1}^L \mathbf{H}_{\beta l}\mathbf{x}_{\beta l}+\mathbf{G}_{\beta}\mathbf{x}_{\alpha}+\mathbf{z}_{\beta}
\end{equation}
where $\mathbf{H}_{\alpha k}\in\mathbb{C}^{N_{\alpha k}\times M_{\alpha}}$, $\mathbf{G}_{\alpha kl}\in\mathbb{C}^{N_{\alpha k}\times N_{\beta l}}$, $\mathbf{H}_{\beta l}\in\mathbb{C}^{M_{\beta}\times N_{\beta l}}$, and $\mathbf{G}_{\beta}\in\mathbb{C}^{M_{\beta}\times M_{\alpha}}$ are the channel matrices from BS $\alpha$ to user $(\alpha, k)$, from user $(\beta, l)$ to user $(\alpha, k)$, from user $(\beta,l)$ to BS $\beta$, and from BS $\alpha$ to BS $\beta$, respectively.
Also, $\mathbf{x}_{\alpha}\in\mathbb{C}^{M_{\alpha}\times 1}$ is the transmit signal vector of BS $\alpha$ and $\mathbf{x}_{\beta l}\in\mathbb{C}^{N_{\beta l}\times 1}$ is the transmit signal vector of user $(\beta, l)$.
The additive noise vectors $\mathbf{z}_{\alpha k}\in\mathbb{C}^{N_{\alpha k}\times 1}$ and $\mathbf{z}_{\beta}\in\mathbb{C}^{M_{\beta}\times 1}$ are assumed to follow $\mathcal{CN}(\mathbf{0}_{N_{\alpha k}\times 1},\mathbf{I}_{N_{\alpha k}})$ and $\mathcal{CN}(\mathbf{0}_{M_{\beta}\times 1},\mathbf{I}_{M_{\beta}})$, respectively.
BS $\alpha$ and each user in cell $\beta$ should satisfy the average power constraint, i.e., $\mathbb{E}(\|\mathbf{x}_{\alpha}\|^2)\leq P_{\alpha}$ and $\mathbb{E}(\|\mathbf{x}_{\beta l}\|^2)\leq P_{\beta l}$ for all $l\in[1:L]$.
We assume that all channel coefficients are independent and identically distributed (i.i.d.) from a continuous distribution. 
Global channel state information is assumed to be available at each user and BS. 

For notational simplicity, denote the considered MIMO R-TDD cellular network as the $\left( {M_\alpha  , \left( {N_{\alpha 1} ,\cdots, N_{\alpha K} } \right)}  \right) \times \left( M_\beta  ,{\left( {N_{\beta 1} ,\cdots, N_{\beta L} } \right)}  \right)$ MIMO R-TDD cellular network.

%%%%%%%%%%%% Sub section %%%%%%%%%%%%
\subsection{Feasibility for Linear Interference Alignment}
Suppose that BS $\alpha$ sends $d_{\alpha k}\in \mathbb{N}$ independent streams, denoted by $\mathbf{s}_{\alpha k}\in \mathbb{C}^{d_{\alpha k}\times 1}$, to user $(\alpha, k)$ using the precoding matrix $\mathbf{V}_{\alpha k}\in\mathbb{C}^{M_{\alpha}\times d_{\alpha k}}$ and user $(\beta, l)$
sends $d_{\beta l}\in \mathbb{N}$ independent streams, denoted by $\mathbf{s}_{\beta l}\in \mathbb{C}^{d_{\beta l}\times 1}$, to BS $\beta$ using the precoding matrix $\mathbf{V}_{\beta l}\in\mathbb{C}^{N_{\beta l}\times d_{\beta l}}$. 
That is, 
\begin{align}
&\mathbf{x}_{\alpha}=\sum_{k=1}^K \mathbf{V}_{\alpha k} \mathbf{s}_{\alpha k},\label{eq:input_alpha} \\ 
&\mathbf{x}_{\beta l}=\mathbf{V}_{\beta l} \mathbf{s}_{\beta l} \label{eq:input_beta}
\end{align}
for $l\in[1:L]$. 
Then user $(\alpha,k)$ estimates $\mathbf{s}_{\alpha k}$ using the postcoding matrix $\mathbf{U}_{\alpha k}\in \mathbb{C}^{N_{\alpha k}\times d_{\alpha k}}$ and BS $\beta$ estimates $\mathbf{s}_{\beta l}$ using the postcoding matrix $\mathbf{U}_{\beta l}\in \mathbb{C}^{M_{\beta}\times d_{\beta l}}$.
That is, 
\begin{align} \label{eq:estimated_alpha}
&\hat{\mathbf{s}}_{\alpha k}\nonumber\\
&=\mathbf{U}_{\alpha k}^{\dagger}\mathbf{y}_{\alpha k}\nonumber\\
&=\sum_{i=1}^K \mathbf{U}_{\alpha k}^{\dagger}\mathbf{H}_{\alpha k}\mathbf{V}_{\alpha i} \mathbf{s}_{\alpha i}+\sum_{l=1}^L \mathbf{U}_{\alpha k}^{\dagger}\mathbf{G}_{\alpha k l}\mathbf{V}_{\beta l} \mathbf{s}_{\beta l}+\mathbf{U}_{\alpha k}^{\dagger}\mathbf{z}_{\alpha k}
\end{align}
for all $k\in[1:K]$ and 
\begin{align}  \label{eq:estimated_beta}
\hat{\mathbf{s}}_{\beta l}&=\mathbf{U}_{\beta l}^{\dagger}\mathbf{y}_{\beta}\nonumber\\
&=\sum_{j=1}^L \mathbf{U}_{\beta l}^{\dagger}\mathbf{H}_{\beta j}\mathbf{V}_{\beta j} \mathbf{s}_{\beta j}+\sum_{k=1}^K \mathbf{U}_{\beta l}^{\dagger}\mathbf{G}_{\beta}\mathbf{V}_{\alpha k} \mathbf{s}_{\alpha k}+\mathbf{U}_{\beta l}^{\dagger} \mathbf{z}_{\beta}
\end{align}
for all $l\in[1:L]$, where the second equality in \eqref{eq:estimated_alpha} follows from \eqref{eq:output_alpha} and \eqref{eq:input_alpha} and the second equality in \eqref{eq:estimated_beta} follows from \eqref{eq:output_beta} and \eqref{eq:input_beta}. %[KY] , -> ?

In similar manners in \cite{Yetis09,Yetis10,Bresler14,Razaviyayn12,Ruan13,Yanjun12,Tingting13,Guillaud11}, from \eqref{eq:estimated_alpha} and \eqref{eq:estimated_beta}, we define the feasible problem of one-shot linear IA for the MIMO R-TDD cellular network as follows.

\begin{definition}[IA feasibility conditions]\label{def:IA_feasibility}%[KY] [IA feasibility] -> [IA feasibility conditions]
 For the MIMO R-TDD cellular network, one-shot linear IA is said to be feasible if there exist $\{\mathbf{U}_{\alpha k},\mathbf{V}_{\alpha k}\}_{k=1}^K$ and $\{\mathbf{U}_{\beta l},\mathbf{V}_{\beta l}\}_{l=1}^L$ satisfying the following set of  
conditions:
\begin{subequations}
\begin{align}
\mathbf{U}_{\alpha k}^{\dagger}\mathbf{G}_{\alpha k l}\mathbf{V}_{\beta l}&=\mathbf{0},{~~}\forall k,l, \label{eq:intercell_IA_alpha}\\
\mathbf{U}_{\beta l}^{\dagger}\mathbf{G}_{\beta}\mathbf{V}_{\alpha k}&=\mathbf{0},{~~}\forall k,l, \label{eq:intercell_IA_beta} \\
\mathbf{U}_{\alpha k}^{\dagger}\mathbf{H}_{\alpha k}\mathbf{V}_{\alpha i}&=\mathbf{0},{~~}\forall i\neq k, \label{eq:intracell_IA_alpha}\\
\mathbf{U}_{\beta l}^{\dagger}\mathbf{H}_{\beta j}\mathbf{V}_{\beta j}&=\mathbf{0},{~~}\forall j\neq l, \label{eq:intracell_IA_beta}\\
{\rm{rank}}\left(\mathbf{U}_{\alpha k}^{\dagger}\mathbf{H}_{\alpha k}\mathbf{V}_{\alpha k}\right)&=d_{\alpha k},{~~}\forall k, \label{eq:desired_alpha}\\
{\rm{rank}}\left(\mathbf{U}_{\beta l}^{\dagger}\mathbf{H}_{\beta l}\mathbf{V}_{\beta l}\right)&=d_{\beta l},{~~}\forall l \label{eq:desired_beta}
\end{align}
\end{subequations}
where $i,k\in[1:K]$ and $j,l\in[1:L]$.\footnote{For notational simplicity, we drop the subscript in $\mathbf{0}_{n\times m}$ when the size of all-zero matrices are clear from the context.}
\end{definition} 

Notice that \eqref{eq:intercell_IA_alpha} to \eqref{eq:intracell_IA_beta} correspond to inter-cell IA from the users in cell $\beta$ to the users in cell $\alpha$, inter-cell IA from BS $\alpha$ to BS $\beta$, intra-cell IA within the users in cell $\alpha$, inter-cell IA within the users in cell $\beta$, respectively.
Assuming the IA feasibility conditions in \eqref{eq:intercell_IA_alpha} to \eqref{eq:intracell_IA_beta}, \eqref{eq:desired_alpha} is required to guarantee that one DoF is able to be delivered by each stream in cell $\alpha$ and \eqref{eq:desired_beta} is required to guarantee that one DoF is able to be delivered by each stream in cell $\beta$. %[KY] within -> among ?

Throughout the paper, we will simply state that $\left( {d_{\alpha 1}},\cdots ,{d_{\alpha K}}, {d_{\beta 1}},\cdots ,{d_{\beta L}} \right)$ is feasible if there is a feasible one-shot linear IA solution satisfying the set of conditions in Definition \ref{def:IA_feasibility} with the given $\left( {d_{\alpha 1}},\cdots ,{d_{\alpha K}}, {d_{\beta 1}},\cdots ,{d_{\beta L}} \right)$.
Obviously, user $(\alpha,k)$ is able to achieve $d_{\alpha k}$ DoF and user $(\beta,l)$ is able to achieve $d_{\beta l}$ DoF in this case.
Denote the feasible sum DoF as
\begin{align}
d_{\sf sum}=  \sum\limits_{k = 1}^K {{d_{\alpha k}}}  + \sum\limits_{l = 1}^L {{d_{\beta l}}}.
\end{align}

In the rest of the paper, we will analyze a necessary condition and a sufficient condition on the IA feasibility for MIMO R-TDD cellular networks described in Definition \ref{def:IA_feasibility}.
For a feasible $d_{\sf sum}$, we state that it is optimal in the following sense.

\begin{definition}[Optimal $d_{\sf sum}$] \label{def:optimal_dof}
The sum DoF $d_{\sf sum}$ is said to be optimal if there exist a necessary condition and a sufficient condition on the IA feasibility in Definition \ref{def:IA_feasibility} such that $d_{\sf sum}$ is the maximum sum DoF not just satisfying the necessary condition but also satisfying the sufficient condition at the same time.
\end{definition}

Note that the optimal $d_{\sf sum}$ in Definition \ref{def:optimal_dof} is the maximum sum DoF achievable by all possible single-shot linear IA strategies.  

\begin{remark}[Frequency division duplex (FDD) systems]
Although we state the IA feasibility based on TDD systems, the main results in this paper also hold for FDD systems.
\end{remark}

  %%%%%%%%%%%% Section %%%%%%%%%%%%
\section{Main Results}\label{sec:main_results}
In this section, we state our main results.
We establish a necessary condition and a sufficient condition on the IA feasibility for MIMO R-TDD cellular networks.
We then provide several example networks that their optimal $d_{\sf sum}$ are characterized by the derived necessary condition and sufficient condition. 
These example networks also demonstrate that the proposed one-shot linear IA using multiple antennas is beneficial for improving the sum DoF of MIMO R-TDD cellular networks.

\begin{theorem}[Necessary condition]\label{the:necessary_conditions}
For the MIMO R-TDD cellular network, any feasible $\left( {d_{\alpha 1}},\cdots ,{d_{\alpha K}}, {d_{\beta 1}},\cdots ,{d_{\beta L}} \right)$ must satisfy the following set of conditions:
\begin{subequations}
\begin{align}
\sum_{k=1}^K d_{\alpha k}&\leq M_{\alpha},\label{eq:necessary(alpha_intra)} \\ 
\sum_{l=1}^L d_{\beta l}&\leq M_{\beta},\label{eq:necessary(beta_intra)}\\
\sum_{k=1}^K d_{\alpha k} + \sum_{l=1}^L d_{\beta l} &\le \max (M_{\alpha} , M_{\beta}),\label{eq:necessary(BS)}\\
\sum\limits_{k \in {\I_{\alpha }}} {{d_{\alpha k}}}  + \sum\limits_{l \in {\I_{\beta}}} {{d_{\beta l}}}  & \le {\rm{max}} \Bigg( {\sum\limits_{k \in {\I_{\alpha}}} {{N_{\alpha k}}}}  , {\sum\limits_{l \in {\I_{\beta}}} {{N_{\beta l}}} } \Bigg), {~~}\forall \I_{\alpha}, \I_{\beta},\label{eq:necessary(user_subset)}\\
%&\!\!\!\!\!\!\!\!\!\!\!\!\!\!\!\!\!\!\!\!\!\!\!\!\!\!\!\!\ 
\sum\limits_{k \in {\I_{\alpha  }}} {\sum\limits_{l \in {\I_{\beta }}} {{d_{\alpha k}}{d_{\beta l}}} } & \le \sum\limits_{k \in {\I_{\alpha  }}} {{d_{\alpha k}}\left( {{N_{\alpha k}} - {d_{\alpha k}}} \right)}  \nonumber \\ 
& {~~}{~~}{~~}{~~}{~}+ \sum\limits_{l \in {\I_{\beta }}} {{d_{\beta l}}\left( {{N_{\beta l}} - {d_{\beta l}}} \right)}, {~~}\forall \I_{\alpha}, \I_{\beta}
\label{eq:necessary(variables_equations)}
\end{align}
\end{subequations}
where ${\I_{\alpha}} \subseteq [1:K]$ and ${\I_{\beta}} \subseteq [1:L]$.
\end{theorem}
\begin{IEEEproof}
We refer to Section \ref{subsec:proof_necessary_conditions} for the proof.
\end{IEEEproof}

In order to state a sufficient condition, partition $\mathbf{G} _{\alpha kl}$ into four sub matrices as 
\begin{align} \label{eq:G_alpha_sub}
\mathbf{G} _{\alpha kl} = \left[ {\begin{array}{*{20}{c}}
{ {\bf{G}} _{\alpha kl}^{(1)}}&{ {\bf{G}} _{\alpha kl}^{(2)}}\\
{ {\bf{G}} _{\alpha kl}^{(3)}}&{ {\bf{G}} _{\alpha kl}^{(4)}}
\end{array}} \right]
\end{align}
where ${\bf{G}} _{\alpha kl}^{(1)} \in \C^{{d_{\alpha k}}  \times {d_{\beta l}} }$, $ {\bf{G}} _{\alpha kl}^{(2)} \in \C^{{d_{\alpha k}}  \times \left( {{N_{\beta l}}  - {d_{\beta l}} } \right)}$, $ {\bf{G}} _{\alpha kl}^{(3)}  \in \C^{\left( {{N_{\alpha k}}  - {d_{\alpha k}} } \right) \times {d_{\beta l}} }$, and $ {\bf{G}} _{\alpha kl}^{(4)}  \in \C^{\left( {{N_{\alpha k}}  - {d_{\alpha k}} } \right) \times \left( {{N_{\beta l}}  - {d_{\beta l}} } \right)} $.
The following theorem establishes a sufficient condition on the IA feasibility.

\begin{theorem}[Sufficient condition]\label{the:sufficient_conditions}
For the MIMO R-TDD cellular network, $\left( {d_{\alpha 1}},\cdots ,{d_{\alpha K}}, {d_{\beta 1}},\cdots ,{d_{\beta L}} \right)$ is feasible almost surely if \eqref{eq:necessary(alpha_intra)} to \eqref{eq:necessary(BS)} are satisfied and \eqref{eq:sufficient IA matrix form}
is a full row rank matrix, where 
 \begin{figure*}[!t]\normalsize
\begin{equation}\label{eq:sufficient IA matrix form}
{{\bf{G}}_\alpha } = \left[ {\begin{array}{*{20}{c}}
{{\bf{G}}_{\alpha 11}^{'}}&{\bf{0}}& \cdots &{\bf{0}}& \cdots &{\bf{0}}&{{\bf{G}}_{\alpha 11}^{''}}&{\bf{0}}& \cdots &{\bf{0}}\\
{{\bf{G}}_{\alpha 12}^{'}}&{\bf{0}}& \cdots &{\bf{0}}& \cdots &{\bf{0}}&{\bf{0}}&{{\bf{G}}_{\alpha 12}^{''}}& \cdots &{\bf{0}}\\
 \vdots & \vdots & \ddots & \vdots & \ddots & \vdots & \vdots & \vdots & \ddots & \vdots \\
{{\bf{G}}_{\alpha 1L}^{'}}&{\bf{0}}& \cdots &{\bf{0}}& \cdots &{\bf{0}}&{\bf{0}}&{\bf{0}}&{\bf{0}}&{{\bf{G}}_{\alpha 1L}^{''}}\\
{\bf{0}}&{{\bf{G}}_{\alpha 21}^{'}}& \cdots &{\bf{0}}& \cdots &{\bf{0}}&{{\bf{G}}_{\alpha 21}^{''}}&{\bf{0}}& \cdots &{\bf{0}}\\
{\bf{0}}&{{\bf{G}}_{\alpha 22}^{'}}& \vdots &{\bf{0}}& \cdots &{\bf{0}}&{\bf{0}}&{{\bf{G}}_{\alpha 22}^{''}}& \cdots &{\bf{0}}\\
 \vdots & \vdots & \ddots & \vdots & \ddots & \vdots & \vdots & \vdots & \ddots & \vdots \\
{\bf{0}}&{{\bf{G}}_{\alpha 2L}^{'}}& \cdots &{\bf{0}}& \cdots &{\bf{0}}&{\bf{0}}&{\bf{0}}& \cdots &{{\bf{G}}_{\alpha 2L}^{''}}\\
 \vdots & \vdots & \vdots & \vdots & \ddots & \vdots & \vdots & \vdots & \vdots & \vdots \\
{\bf{0}}&{\bf{0}}& \cdots &{\bf{0}}& \cdots &{{\bf{G}}_{\alpha K1}^{'}}&{{\bf{G}}_{\alpha K1}^{''}}&{\bf{0}}& \cdots &{\bf{0}}\\
{\bf{0}}&{\bf{0}}& \cdots &{\bf{0}}& \cdots &{{\bf{G}}_{\alpha K2}^{'}}&{\bf{0}}&{{\bf{G}}_{\alpha K2}^{''}}& \cdots &{\bf{0}}\\
 \vdots & \vdots & \ddots & \vdots & \ddots & \vdots & \vdots & \vdots & \ddots & \vdots \\
{\bf{0}}&{\bf{0}}& \cdots &{\bf{0}}& \cdots &{{\bf{G}}_{\alpha KL}^{'}}&{\bf{0}}&{\bf{0}}& \cdots &{{\bf{G}}_{\alpha KL}^{''}}
\end{array}} \right]\end{equation}  
\hrulefill
\vspace*{4pt}
\end{figure*}
\setlength{\arraycolsep}{0.14em}
\begin{eqnarray}
\mathbf{G}^{'}_{\alpha k l}&=&\operatorname{diag}[d_{\alpha k}]\left(\mathbf{G}^{(3)T}_{\alpha k l}\right),\nonumber\\
\mathbf{G}^{''}_{\alpha k l}&=& \left[ {\begin{array}{*{20}{c}}
{{\rm{diag}}\left[ {{d_{\beta l}}} \right]\left( { {\bf{G}} _{\alpha kl}^{(2)}\left[ 1 \right]} \right)}\\
{{\rm{diag}}\left[ {{d_{\beta l}}} \right]\left( { {\bf{G}} _{\alpha kl}^{(2)}\left[ 2 \right]} \right)}\\
 \vdots \\
{{\rm{diag}}\left[ {{d_{\beta l}}} \right]\left( { {\bf{G}} _{\alpha kl}^{(2)}\left[ {{d_{\alpha k}}} \right]} \right)}
\end{array}} \right]\nonumber
\end{eqnarray}
and the definitions of $\mathbf{G}^{(2)}_{\alpha k l}$ and $\mathbf{G}^{(3)}_{\alpha k l}$ are given by \eqref{eq:G_alpha_sub}.
\end{theorem}
\begin{IEEEproof}
We refer to Section \ref{subsec:proof_sufficient_conditions} for the proof.
\end{IEEEproof}

Depending on the network configuration, there may exist a DoF gap between the necessary condition and the sufficient condition stated in Theorems \ref{the:necessary_conditions} and \ref{the:sufficient_conditions} respectively.
The following two examples show one case where the sum DoF satisfying the necessary condition and the sufficient condition are the same, thereby it is optimal from Definition \ref{def:optimal_dof}, and the other case where there exist a DoF gap.
For both cases, more importantly, the proposed one-shot linear IA strictly enlarges the sum DoF compared with the sum DoF achievable by operating one of the two cells, which is given by
\begin{align} \label{eq:single_cell_lower}
d_{\sf sum, single}=\max\Bigg\{&\min\left(M_{\alpha},\sum_{k=1}^KN_{\alpha k}\right),
\nonumber\\
&{~~}{~~}{~~}{~~}{~~}\min\left(M_{\beta},\sum_{l=1}^L N_{\beta l} \right) \Bigg\}.
\end{align}
%da = 2 4 4, db = 1 2

\begin{example}[An example network where its optimal $d_{\sf sum}$ is characterized by Theorems \ref{the:necessary_conditions} and \ref{the:sufficient_conditions}]\label{ex:necessary_sufficient}
Consider the $\left( 10,(4,6,6) \right) \times \left( 13,(3,6) \right)$ MIMO R-TDD cellular network. 
For this configuration, Theorem \ref{the:necessary_conditions} implies $d_{\sf sum}\leq 13$, which is feasible from Theorem \ref{the:sufficient_conditions}. Therefore, the optimal sum DoF is given by $d_{\sf sum}=13$.
However, the single-cell lower bound in \eqref{eq:single_cell_lower} only achieves $d_{\sf sum, single} = 10$.
\end{example}

\begin{example}[An example network where there is a DoF from Theorems \ref{the:necessary_conditions} and \ref{the:sufficient_conditions}]\label{rem:gap_necessary_sufficeint}
Consider the $\left( 8,(2,3,8) \right) \times \left( 12,(3,7) \right)$ MIMO R-TDD cellular network.
For this configuration, Theorem \ref{the:necessary_conditions} implies $d_{\sf sum}\leq 12$ but $d_{\sf sum}= 11$ is feasible from Theorem \ref{the:sufficient_conditions}, which shows the sum DoF gap of one.
Although there exists the DoF gap from its upper bound, the proposed IA in Theorem \ref{the:sufficient_conditions} strictly improves the sum DoF compared with the single-cell lower bound, given by $d_{\sf sum, single}=10$.
\end{example}

\begin{remark}[Duality for the IA feasibility]\label{remark:duality property} 
Suppose that a DoF tuple $\left( {d_{\alpha 1}},\cdots ,{d_{\alpha K}}, {d_{\beta 1}},\cdots ,{d_{\beta L}} \right)$ satisfies the set of conditions in  Theorem \ref{the:sufficient_conditions}, meaning that it is feasible, for the $\left( {M_\alpha  , \left( {N_{\alpha 1} ,\cdots, N_{\alpha K} } \right)}  \right) \times \left( M_\beta  ,{\left( {N_{\beta 1} ,\cdots, N_{\beta L} } \right)}  \right)$ MIMO R-TDD cellular network.
Then $\left( {d_{\beta 1}},\cdots ,{d_{\beta L}}, {d_{\alpha 1}},\cdots ,{d_{\alpha K}} \right)$ is feasible for its dual MIMO R-TDD cellular network, i.e., the $\left( M_\beta  ,{\left( {N_{\beta 1} ,\cdots, N_{\beta L} } \right)}  \right) \times \left( {M_\alpha  , \left( {N_{\alpha 1} ,\cdots, N_{\alpha K} } \right)}  \right)$  MIMO R-TDD cellular network.
\end{remark}

\begin{example}[The dual network of Example \ref{ex:necessary_sufficient}]
From Example \ref{ex:necessary_sufficient}, the optimal sum DoF of the $\left( 13,(3,6) \right) \times \left( 10,(4,6,6) \right)$ MIMO R-TDD cellular network is also given by $d_{\sf sum} = 13$.
\end{example} 

In many cases of interest, each user may require the same DoF, i.e., ${d_{\alpha k}}={d_\alpha}, {d_{\beta l}}={d_\beta}$ for all $k\in[1:K]$ and $l\in[1:L]$.
By focusing on such a symmetric DoF, we establish a sufficient condition with a more explicit expression than Theorem \ref{the:sufficient_conditions}.

\begin{theorem}[Sufficient condition for symmetric DoF]\label{the:divisible}

For the MIMO R-TDD cellular network, a symmetric DoF $d_\alpha$ and $d_\beta$, i.e., ${d_{\alpha k}}={d_\alpha}, {d_{\beta l}}={d_\beta}$ for all $k\in[1:K]$ and $l\in[1:L]$, is feasible almost surely if the following set of conditions are satisfied:
\begin{subequations}
\begin{align}
Kd_{\alpha} & \leq M_{\alpha},\label{eq:cor1(alpha_intra)} \\ 
Ld_{\beta} & \leq M_{\beta},\label{eq:cor1(beta_intra)}\\
Kd_{\alpha} + Ld_{\beta} &\le \max (M_{\alpha} , M_{\beta}),\label{eq:cor1(BS)}\\
{\rm{mod}}(N_{\alpha k}-d_\alpha ,d_\beta)& = {\rm{mod}}(N_{\beta l}-d_\beta,d_\alpha)=0, {~~}\forall {k,l}, \label{eq:cor1(divisible)}\\
\left| {\I_\alpha}\right| \left|{\I_\beta} \right| d_\alpha d_\beta & \le \sum\limits_{k \in {\I_{\alpha  }}} {d_\alpha \left( {{N_{\alpha k}} - d_\alpha} \right)} \nonumber \\
&{~~~~~~}+ \sum\limits_{l \in {\I_{\beta }}} {{d_\beta}\left( {{N_{\beta l}} - {d_\beta}} \right)}, {~~}\forall \I_{\alpha}, \I_{\beta}
\label{eq:cor1(variables_equations)}
\end{align}
\end{subequations}
where $k\in[1:K]$, $l\in[1:L]$, ${\I_{\alpha}} \subseteq [1:K]$, and ${\I_{\beta}} \subseteq [1:L]$.
\end{theorem}
\begin{IEEEproof}
See Section \ref{subsec:proof_cor1} for the proof.
\end{IEEEproof}

\begin{remark}[Necessary and sufficient condition for divisible symmetric DoF] For symmetric DoF, the conditions \eqref{eq:necessary(alpha_intra)}, \eqref{eq:necessary(beta_intra)}, \eqref{eq:necessary(BS)}, and \eqref{eq:necessary(variables_equations)} in Theorem \ref{the:necessary_conditions} yield the conditions \eqref{eq:cor1(alpha_intra)}, \eqref{eq:cor1(beta_intra)}, \eqref{eq:cor1(BS)}, and \eqref{eq:cor1(variables_equations)} in Theorem \ref{the:divisible}. Therefore, if \eqref{eq:cor1(divisible)} is satisfied for given $d_\alpha$ and $d_\beta$, then Theorem 3 provides the necessary and sufficient condition, i.e., the set of conditions \eqref{eq:cor1(alpha_intra)}, \eqref{eq:cor1(beta_intra)}, \eqref{eq:cor1(BS)}, and \eqref{eq:cor1(variables_equations)}.
\end{remark}

\begin{example}[An example network with symmetric DoF]\label{ex:symmetric}
Consider the $\left( 12,(6,6,8) \right) \times \left( 16,(6,6) \right)$ MIMO R-TDD cellular network.
Then $d_\alpha =4$ and $d_\beta = 2$ are feasible from Theorem \ref{the:divisible}, which achieves $d_{\sf sum} = 16$.
Theorem \ref{the:necessary_conditions} implies $d_{\sf sum} \le 16$ for this configuration, hence, it is the optimal sum DoF and provides a larger sum DoF than $d_{\sf sum, single} = 12$. % [KY] optimal -> the optimal
\end{example}

\section{IA Feasibility for MIMO R-TDD cellular Networks} \label{sec:proof_of_results}
In this section, we prove Theorems \ref{the:necessary_conditions}, \ref{the:sufficient_conditions}, and \ref{the:divisible} stated in Section \ref{sec:main_results}.

\begin{figure*}[!t]
\centering
\subfloat[$M_\alpha \ge M_\beta$]{\includegraphics[width=3.5in]{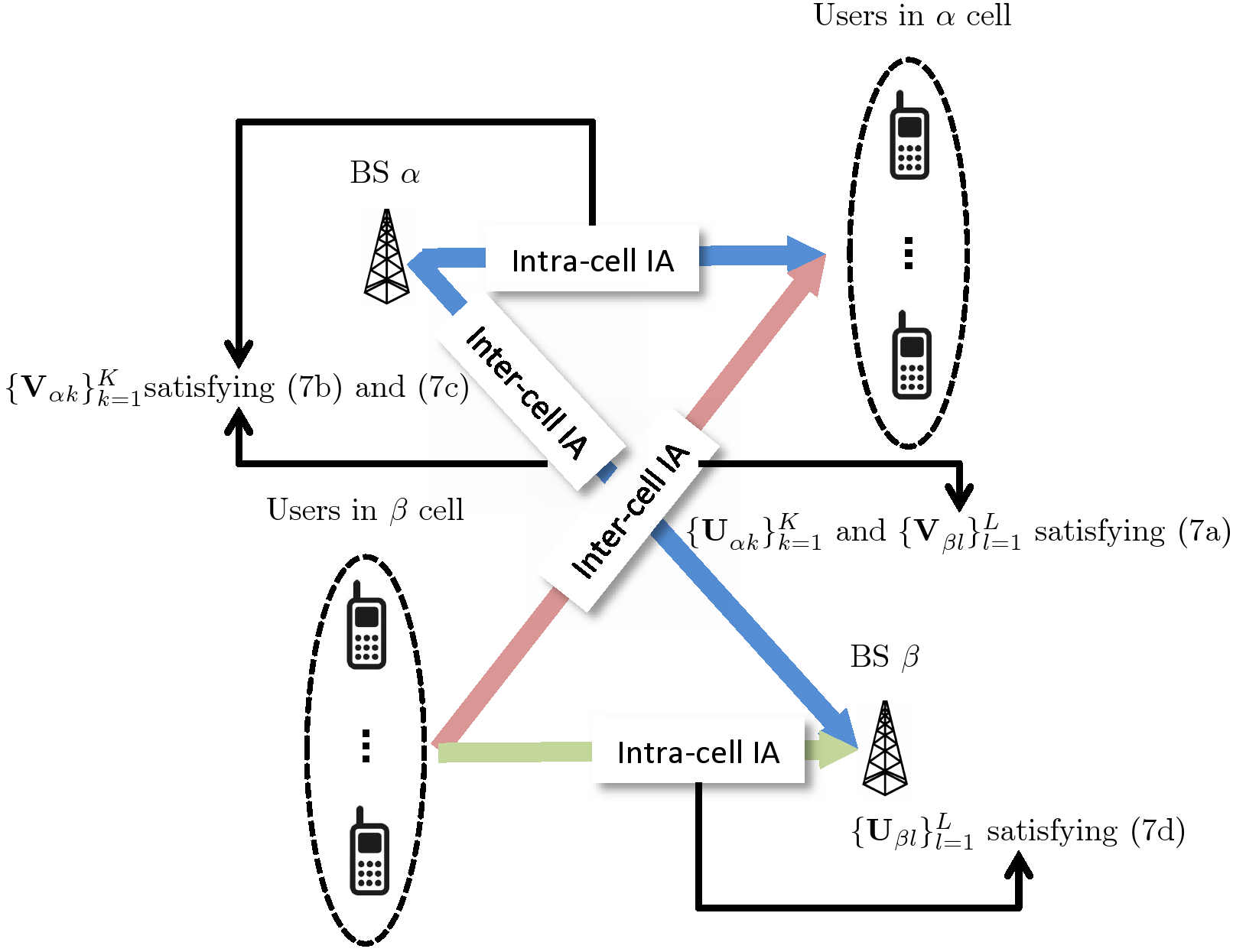}
\label{fig:sub1}}
\hfil
\subfloat[$M_\alpha < M_\beta$]{\includegraphics[width=3.5in]{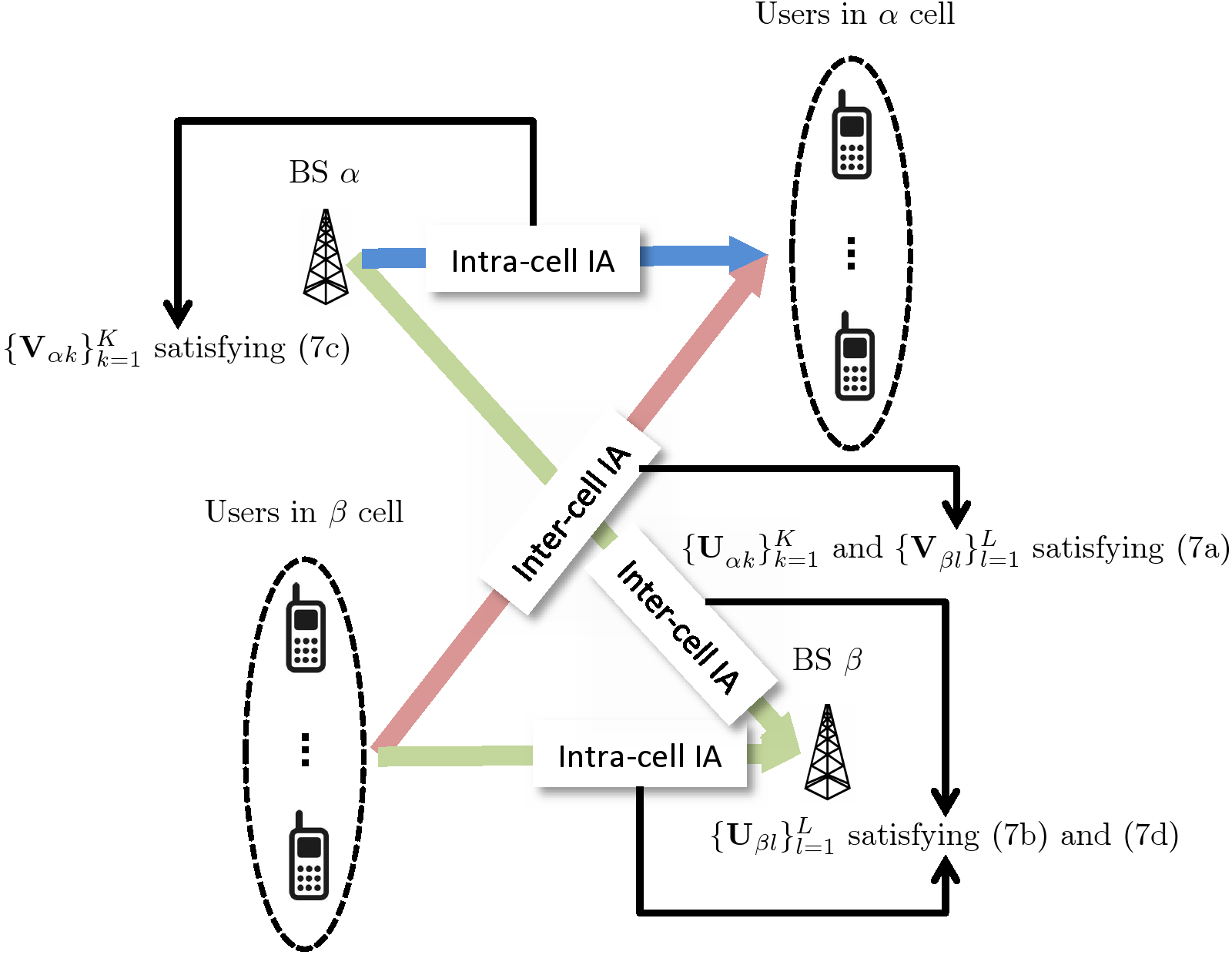}
\label{fig:sub2}}
\caption{The relations of the precoding, postcoding matrices and its interferences.}
\label{fig:building_concept}
\end{figure*}

\subsection{Proof of Theorem \ref{the:necessary_conditions}}\label{subsec:proof_necessary_conditions}
In order to prove Theorem \ref{the:necessary_conditions}, we first introduce the sum DoF of the two-user MIMO IC in \cite{Jafar07}. For the two-user MIMO IC with  $M_1$ transmit antennas and $N_1$ received antennas for the first transmission pair and $M_2$ transmit antennas and $N_2$ received antennas for the second transmission pair, the optimal sum DoF is given by 
\begin{equation} \label{eq:dof_mimo_ic}
\min\left\{M_1+M_2, N_1+N_2,\max (M_1,N_2),\max(M_2,N_1)\right\}.
\end{equation}
From \eqref{eq:dof_mimo_ic}, for given $\mathcal{I}_{\alpha }\subseteq[1:K]$ and $\mathcal{I}_{\beta }\subseteq[1:L]$,
\begin{align}\label{eq:necessary_conditions(total)}
&\sum_{k\in\mathcal{I}_{\alpha }}d_{\alpha k}+\sum_{l\in\mathcal{I}_{\beta }}d_{\beta l} \nonumber \\
& \leq\min \Bigg\{ M_{\alpha}+\sum_{l\in\mathcal{I}_{\beta }}N_{\beta l},M_{\beta}+\sum_{k\in\mathcal{I}_{\alpha  }}N_{\alpha k},\nonumber \\ 
&{~~~~~~~~~} \max(M_{\alpha},M_{\beta}),\max \bigg( \sum_{k\in\mathcal{I}_{\alpha }}N_{\alpha k},\sum_{l\in\mathcal{I}_{\beta}}N_{\beta l} \bigg) \Bigg\}
\end{align}
which corresponds to the sum DoF if full cooperation is allowed among the users in $\mathcal{I}_{\alpha}$ and the users in $\mathcal{I}_{\beta}$ respectively. %[KY] between -> among  &   ~~~~ respectively.
Hence \eqref{eq:necessary_conditions(total)} must be satisfied for all $\mathcal{I}_{\alpha }\subseteq[1:K]$ and $\mathcal{I}_{\beta }\subseteq[1:L]$, which yields the conditions \eqref{eq:necessary(alpha_intra)} to \eqref{eq:necessary(user_subset)}.

The last condition \eqref{eq:necessary(variables_equations)} is derived in a similar manner in \cite{Razaviyayn12}.
From the IA feasibility conditions \eqref{eq:desired_alpha} and \eqref{eq:desired_beta}, $\{ {{\mathbf{U}}_{\alpha k}}\}_{k=1}^{K}$ and $\{ {{\bf{V}}_{\beta l}} \}_{l=1}^{L}$ should be full column rank matrices. %[KY] IA conditions -> IA feasibility conditions
Therefore, we restrict ${{\mathbf{U}}_{\alpha k}}$ and ${{\bf{V}}_{\beta l}}$ such that  ${\rm{rank}}\left( {{\mathbf{U}}_{\alpha k}} \right) = {d_{\alpha k}} $ and ${\rm{rank}}\left(  {{\bf{V}}_{\beta l}}  \right) =  {d_{\beta l}} $ for all $k\in[1:K]$ and $l\in[1:L]$. %[KY] -> for all $k\in[1:K]$ and $l\in[1:L]$.
Then they can be rewritten as 
\begin{equation}\label{eq:rearrange coders}
{{\bf{U}}_{\alpha k}} = \left[ {\begin{array}{*{20}{c}}
{{{\bf{I}}_{{d_{\alpha k}}}}}\\
{{{\overline {\bf{U}} }_{\alpha k}}}
\end{array}} \right]{\bf{A}}_{\alpha k}^{ - 1},{{\bf{V}}_{\beta l}} = \left[ {\begin{array}{*{20}{c}}
{{{\bf{I}}_{{d_{\beta l}}}}}\\
{{{\overline {\bf{V}} }_{\beta l}}}
\end{array}} \right]{\bf{A}}_{\beta l}^{ - 1}
\end{equation}
where ${\mathbf{\overline U}}_{\alpha k}  \in \C^{\left( {{N_{\alpha k}}  - {d_{\alpha k}} } \right) \times {d_{\alpha k}} }$ and ${\mathbf{\overline V}}_{\beta l}  \in \C^{\left( {{N_{\beta l}}  - {d_{\beta l}} } \right) \times {d_{\beta l}} }$ are some arbitrary matrices and ${{\bf{A}}_{\alpha k}}  \in \C^{{N_{\alpha k}}  \times {N_{\alpha k}} }$ and ${{\bf{A}}_{\beta l}}  \in \C^{{N_{\beta l}}  \times {N_{\beta l}} } $ are some invertible matrices. 

From \eqref{eq:G_alpha_sub}  and \eqref{eq:rearrange coders}, the IA feasibility condition \eqref{eq:intercell_IA_alpha} is given as % [KY] IA -> IA feasibility
\begin{equation}\label{eq:Reform_7a}
\left[ {\begin{array}{*{20}{c}}
{{{\bf{I}}_{{d_{\alpha k}}}}}&{\overline {\bf{U}} _{\alpha k}^\dag }
\end{array}} \right]\left[ {\begin{array}{*{20}{c}}
{{\bf{G}}_{\alpha kl}^{(1)}}&{{\bf{G}}_{\alpha kl}^{(2)}}\\
{{\bf{G}}_{\alpha kl}^{(3)}}&{{\bf{G}}_{\alpha kl}^{(4)}}
\end{array}} \right]\left[ {\begin{array}{*{20}{c}}
{{{\bf{I}}_{{d_{\beta l}}}}}\\
{{{\overline {\bf{V}} }_{\beta l}}}
\end{array}} \right] = {\bf{0}}
\end{equation}
for all $k \in [1:K]$ and $l \in [1:L]$.
Let us now consider fixed $\mathcal{I}_{\alpha}$ and $\mathcal{I}_{\beta}$, where $\mathcal{I}_{\alpha}\subseteq[1:K]$ and $\mathcal{I}_{\beta}\subseteq[1:L]$.
By treating the elements of ${\mathbf{\overline U}}_{\alpha k}$ and ${\mathbf{\overline V}}_{\beta l}$ as controllable variables and considering a set of conditions for all $k\in \mathcal{I}_{\alpha}$ and $l\in\mathcal{I}_{\beta}$ in \eqref{eq:Reform_7a}, we have
\begin{align*}
\sum\limits_{k \in \mathcal{I}_\alpha } {\sum\limits_{l \in \mathcal{I}_\beta} {{d_{\alpha k}} {d_{\beta l}} } }  &\leq \sum\limits_{k \in \mathcal{I}_\alpha} {d_{\alpha k}}{\left( {{N_{\alpha k}}  - {d_{\alpha k}} } \right)   } \nonumber\\
&{~~~~~~~~~~~~~~~~~~} + \sum\limits_{l \in \mathcal{I}_\beta} {d_{\beta l}}{\left( {{N_{\beta l}}  - {d_{\beta l}} } \right) }
\end{align*}
where the left-hand side corresponds to the number of equations and the right-hand side  corresponds to the number of variables.
Since the above inequality should be satisfied for all $\mathcal{I}_{\alpha}\subseteq[1:K]$ and $\mathcal{I}_{\beta}\subseteq[1:L]$, we have the condition \eqref{eq:necessary(variables_equations)}.
In conclusion, Theorem \ref{the:necessary_conditions} holds.

\subsection{Proof of Theorem \ref{the:sufficient_conditions}}\label{subsec:proof_sufficient_conditions}
In this subsection, we prove Theorem \ref{the:sufficient_conditions}.
We show that precoding and postcoding matrices satisfying the IA feasibility conditions in Definition \ref{def:IA_feasibility} exist almost surely if the set of conditions in Theorem \ref{the:sufficient_conditions} are satisfied. The overall construction of precoding and postcoding matrices is as follows: %[KY] IA feasibility -> IA feasibility conditions
\begin{itemize}  
\item Step 1:  
Construct $\{{\bf{U}}_{\alpha k}\}_{k=1}^K$ and $\{{\bf{V}}_{\beta l}\}_{l=1}^L$ satisfying \eqref{eq:intercell_IA_alpha} as shown in both Fig. \ref{fig:sub1} and Fig. \ref{fig:sub2}. %[KY]+ as shown in both Fig. \ref{fig:sub1} and \ref{fig:sub2}.
\item Step 2:
For $ M_{\alpha}\geq M_{\beta}$, first construct $\{\mathbf{U}_{\beta l}\}_{l=1}^L$  satisfying \eqref{eq:intracell_IA_beta} and then construct $\{\mathbf{V}_{\alpha k}\}_{k=1}^K$ satisfying \eqref{eq:intercell_IA_beta} and \eqref{eq:intracell_IA_alpha} as shown in Fig. \ref{fig:sub1}.
On the other hand, for $ M_{\alpha} < M_{\beta}$, first construct $\{\mathbf{V}_{\alpha k}\}_{k=1}^K$ satisfying \eqref{eq:intracell_IA_alpha} and then construct $\{\mathbf{U}_{\beta l}\}_{l=1}^L$ satisfying \eqref{eq:intercell_IA_beta} and \eqref{eq:intracell_IA_beta} as shown in Fig. \ref{fig:sub2}.
\end{itemize}

In the following, we show that $\{{\bf{U}}_{\alpha k}\}_{k=1}^K$ and $\{{\bf{V}}_{\beta l}\}_{l=1}^L$ in Step 1 can be established almost surely if $\mathbf{G}_{\alpha}$ in \eqref{eq:sufficient IA matrix form} is a full row rank matrix and $\{\mathbf{U}_{\beta l}\}_{l=1}^L$ and $\{\mathbf{V}_{\alpha k}\}_{k=1}^K$  in Step 2 can be established almost surely if the conditions in \eqref{eq:necessary(alpha_intra)} to \eqref{eq:necessary(BS)} are satisfied.
In Section \ref{sec:IAbuilder}, we further explain in details how to construct such $\{ {\mathbf{U}}_{\alpha k}, {\mathbf{V}}_{\alpha k} \}_{k=1}^K$ and $\{ {\mathbf{U}}_{\beta l},{\mathbf{V}}_{\beta l} \}_{l=1}^L$. %[KY]
Especially, $\{ {\mathbf{U}}_{\alpha k} \}_{k=1}^K$ and $\{ {\mathbf{V}}_{\beta l} \}_{l=1}^L$ are constructed by using the iterative algorithm proposed in \cite{Gomadam11}. %[KY]+  adding a new sentence

\begin{remark}[Inter-cell IA from BS $\alpha$ to BS $\beta$] For the above construction, inter-cell IA is applied at BS $\alpha$ using precoding matrices in order to satisfy  \eqref{eq:intercell_IA_beta} and then BS $\beta$ decodes its streams using postcoding matrices as described in Fig. \ref{fig:sub1} if $M_{\alpha}\geq M_{\beta}$. %[KY] matrices if -> matrices as described in Fig. \ref{fig:sub1} if
On the other hand, BS $\beta$ decodes its streams without inter-cell IA at BS $\alpha$ as described in Fig. \ref{fig:sub2} if $M_{\alpha}< M_{\beta}$. %[KY] at BS $\alpha$ if -> at BS $\alpha$ as described in Fig. \ref{fig:sub2} if
It has been shown in \cite{Jeon14arxiv} that this approach achieves the optimal sum DoF for the case where each user has a single antenna.
\end{remark}

\subsubsection{Construction of $\{{\bf{U}}_{\alpha k}\}_{k=1}^K$ and $\{{\bf{V}}_{\beta l}\}_{l=1}^L$}
We first concentrate on constructing $\{{\bf{U}}_{\alpha k}\}_{k=1}^K$ and $\{{\bf{V}}_{\beta l}\}_{l=1}^L$ that satisfies \eqref{eq:intercell_IA_alpha}.
Since these matrices should also satisfy \eqref{eq:desired_alpha} and \eqref{eq:desired_beta}, we assume that 
\begin{equation}\label{eq:U_V_}
{{\bf{U}}_{\alpha k}}=\left[ {\begin{array}{*{20}{c}}
{{{\bf{I}}_{{d_{\alpha k}}}}}\\
{{{\overline {\bf{U}} }_{\alpha k}}}
\end{array}} \right] ,{{\bf{V}}_{\beta l}}=\left[ {\begin{array}{*{20}{c}}
{{{\bf{I}}_{{d_{\beta l}}}}}\\
{{{\overline {\bf{V}} }_{\beta l}}}
\end{array}} \right].
\end{equation}
That is, we only control ${\mathbf{\overline U}}_{\alpha k}  \in \C^{\left( {{N_{\alpha k}}  - {d_{\alpha k}} } \right) \times {d_{\alpha k}} }$ and ${\mathbf{\overline V}}_{\beta l}  \in \C^{\left( {{N_{\beta l}}  - {d_{\beta l}} } \right) \times {d_{\beta l}} }$ in order to satisfy \eqref{eq:intercell_IA_alpha}.
Let
\begin{align}
&q\left( {k,l,m,n} \right) \nonumber \\
&{~~}=\sum\limits_{k' = 1}^{k - 1} {\sum\limits_{l' = 1}^l {d_{\alpha k'} d_{\beta l'} } }  + \sum\limits_{l' = 1}^l {d_{\alpha k} d_{\beta l'} } + \left( {m - 1} \right)d_{\beta l'}  + n \nonumber
\end{align}
for $m\in[1:d_{\alpha k}]$ and $n\in[1:d_{\beta l}]$.
Then define $r_{q(k,l,m,n)}$ as the $(m,n)$th element of $\mathbf{U}_{\alpha k}^{\dagger}\mathbf{G}_{\alpha k l}\mathbf{V}_{\beta l}$, that is given by
\begin{align}\label{eq:IA received signal}
&r_{q(k,l,m,n)} \nonumber \\
&{~~}={\bf G} _{\alpha kl}^{(1)}[ {m,n} ] + 
\sum\limits_{i = 1}^{{N_{\beta l}} - {d_{\beta l}}} {{\bf G} _{\alpha kl}^{(2)} [ {m,i} ]{{\overline {\bf V} }_{\beta l}}[ {i,n} ]} \nonumber \\
&{~~~~~} + \sum\limits_{j = 1}^{{N_{\alpha k}} - {d_{\alpha k}}} { {\bf G} _{\alpha kl}^{(3)}[ {j,m} ] \overline {\bf U} _{\alpha k}^\dag [ {j,n} ]} \nonumber\\
&{~~~~~}+ \sum\limits_{j = 1}^{{N_{\alpha k}} - {d_{\alpha k}}} {\sum\limits_{i = 1}^{{N_{\beta l}} - {d_{\beta l}}} {{\bf G} _{\alpha kl}^{(4)}[ {j,i} ] \overline {\bf U} _{\alpha k}^\dag [ {j,n} ] {{\overline {\bf V} }_{\beta l}} [ {i,n} ]} } \nonumber\\
&{~~} = {\bf G} _{\alpha kl}^{(1)}[ {m,n} ] + {\bf{G}} _{{\alpha}} [ q( {k,l,m,n} ) ] {\bf{f}}_1 \nonumber \\
&{~~~~~}+ \sum\limits_{j = 1}^{{N_{\alpha k}} - {d_{\alpha k}}} {\sum\limits_{i = 1}^{{N_{\beta l}} - {d_{\beta l}}} {{\bf G} _{\alpha kl}^{(4)}[ {j,i} ]\overline {\bf U} _{\alpha k}^\dag [ {j,n} ] {{\overline {\bf V} }_{\beta l}} [ {i,n} ] } }
\end{align}
where 
\begin{equation}\label{eq:IA variables}
{\bf{f}}_1 =
\begin{bmatrix}
{\rm{vec}}{{\left( {\overline {\bf{U}} _{\alpha 1}^\dag } \right)}}\\ 
\vdots \\
{\rm{vec}}{{\left( {\overline {\bf{U}} _{\alpha K}^\dag } \right)}}\\
{\rm{vec}}{{\left( {{{\overline {\bf{V}} }_{\beta 1}}} \right)}}\\
\vdots\\
{\rm{vec}}{{\left( {{{\overline {\bf{V}} }_{\beta L}}} \right)}} 
\end{bmatrix}.
\end{equation}
Hence, from \eqref{eq:IA received signal}, \eqref{eq:intercell_IA_alpha} can be expressed as 
\begin{equation}\label{eq:cond4}
\begin{bmatrix}
r_1\\
r_2\\
\vdots\\
r_{\sum_{k'=1}^K\sum_{l'=1}^L d_{\alpha k'} d_{\beta l'}}
\end{bmatrix} =\mathbf{f}_0+\mathbf{G}_{\alpha}\mathbf{f}_1+\mathbf{f}_2=\mathbf{0}
\end{equation}
where $\mathbf{f}_0$ is the zero-order polynomial vector, independent of $\{{\mathbf{\overline U}}_{\alpha k}\}_{k=1}^K$ and $\{{\mathbf{\overline V}}_{\beta l}\}_{l=1}^L$, and $\mathbf{G}_{\alpha}\mathbf{f}_1$ and $\mathbf{f}_2$ are respectively the first-order and the second-order polynomial vectors with respect to  $\{{\mathbf{\overline U}}_{\alpha k}\}_{k=1}^K$ and $\{{\mathbf{\overline V}}_{\beta l}\}_{l=1}^L$. %[KY]  second-order ->  the second-order
The definition of $\mathbf{G}_{\alpha}$ is given by \eqref{eq:sufficient IA matrix form}.

In order to show the existence of $\{{\mathbf{\overline U}}_{\alpha k}\}_{k=1}^K$ and $\{{\mathbf{\overline V}}_{\beta l}\}_{l=1}^L$ satisfying \eqref{eq:cond4}, we introduce the following algebraic geometric lemma in \cite{Ruan13}, which has been used for the feasibility problem of the MIMO $K$-user IC. %[KY] in \cite{book:knapp07}, which has been used for the feasibility problem of the MIMO $K$-user IC in \cite{Ruan13}. -> in \cite{Ruan13}, which has been used for the feasibility problem of the MIMO $K$-user IC.

%\begin{lemma}[Linear independence leads to a solution]\label{lem:linear_solution}
%Consider a set of $R$ polynomials represented by $\{g_i \in {\mathbb {C}}[x_1 , x_2 , \cdots ,x_S]\}_{i \in [1:R]}$, where $g_i = \sum_{j=1}^S f_{ij} x_j + e_i$ and $e_i$ is a set of polynomials with degrees no less than two. 
%If the coefficient vectors $\{ \mathbf{g}_i = [f_{i1}, f_{i2}, \cdots , f_{iS}]\}_{i \in [1:R]}$ are linearly independent, then the polynomials $\{g_i\}_{i \in [1:R]}$ are algebraically independent and, as a result, the linear equations $\{g_i = c_i\}_{i \in [1:R]}$ has a solution almost sourly, where the elements of $\{c_i\in\mathbb{C}\}_{i \in [1:R]}$ are independently drawn from continuous distributions.
%\end{lemma} % original

\begin{lemma}[Linear independence leads to a solution]\label{lem:linear_solution}
Consider a set of $R$ polynomials represented by $\{g_i \in {\mathbb {C}}[x_1 , x_2 , \cdots ,x_S]\}_{i=1}^{R}$, where ${\mathbb {C}}[x_1 , x_2 , \cdots ,x_S]$ denotes an algebraically closed field of rational functions in variables $x_1 , x_2 , \cdots ,x_S$ with coefficients drawn from $\C$, $g_i = \sum_{j=1}^S f_{ij} x_j + e_i$, and $e_i$ is a set of polynomials with degrees no less than two.
If the coefficient vectors $\mathbf{g}_i =[ {\begin{array}{*{20}{c}} {f_{i1}}& {f_{i2}}&\cdots &{f_{iS}} \end{array}} ]_{i=1}^{R}$ are linearly independent, then the polynomials $\{g_i\}_{i=1}^{R}$ are algebraically independent and, as a result, the linear equations $\{g_i = c_i\}_{i=1}^{R}$ have a solution almost surely, where the elements of $\{c_i\in\mathbb{C}\}_{i=1}^{R}$ are independently drawn from continuous distributions. 
\end{lemma} %[KY] _{i \in [1:R]} -> _{i=1}^{R} and add up a notation of field

Notice that the above lemma can be applied to $\eqref{eq:cond4}$ since the elements of $\mathbf{f}_0$ are independently drawn from continuous distributions.
Therefore, $\{{\mathbf{\overline U}}_{\alpha k}\}_{k=1}^K$ and $\{{\mathbf{\overline V}}_{\beta l}\}_{l=1}^L$ satisfying \eqref{eq:cond4} exist almost surely if the row vectors of $\mathbf{G}_{\alpha}$ are linearly independent.
In conclusion, $\{ {\mathbf{U}}_{\alpha k} \}_{k=1}^K$ and $\{ {\mathbf{V}}_{\beta l} \}_{l=1}^L$ satisfying \eqref{eq:intercell_IA_alpha} exist almost surely if $\mathbf{G}_{\alpha}$ is a full row rank matrix.
%In Section \ref{sec:IAbuilder}, we explain in details how to construct such $\{ {\mathbf{U}}_{\alpha k} \}_{k=1}^K$ and $\{ {\mathbf{V}}_{\beta l} \}_{l=1}^L$ using the iterative algorithm proposed in  \cite{Gomadam11}.

\subsubsection{Construction of $\{\mathbf{V}_{\alpha k}\}_{k=1}^K$ and $\{\mathbf{U}_{\beta l}\}_{l=1}^L$}
Let us now consider the construction of $\{\mathbf{V}_{\alpha k}\}_{k=1}^K$ and $\{\mathbf{U}_{\beta l}\}_{l=1}^L$, for given $\{{\bf{U}}_{\alpha k}\}_{k=1}^K$ and $\{{\bf{V}}_{\beta l}\}_{l=1}^L$ satisfying \eqref{eq:intercell_IA_alpha}.

First consider the case where $ M_{\alpha}\geq M_{\beta}$.
For this case, the condition \eqref{eq:necessary(alpha_intra)} becomes inactive from the condtion \eqref{eq:necessary(BS)}.
We construct $\{\mathbf{U}_{\beta l}\}_{l=1}^L$ only satisfying \eqref{eq:intracell_IA_beta}.
Then there exists a nonzero $\mathbf{U}_{\beta l}$ satisfying \eqref{eq:intracell_IA_beta} almost surely if 
\begin{equation*}
d_{\beta l}\leq M_{\beta}-\sum_{l'=1,l'\neq l}^L d_{\beta l'}
\end{equation*}
since the total number of intra-cell interference dimensions is given by $\sum_{l'=1,l'\neq l}^L d_{\beta l'}$, which yields the condition \eqref{eq:necessary(beta_intra)}.
Then we construct $\{\mathbf{V}_{\alpha k}\}_{k=1}^K$ satisfying \eqref{eq:intercell_IA_beta} and \eqref{eq:intracell_IA_alpha} for given $\{{\bf{U}}_{\alpha k}\}_{k=1}^K$ and $\{\mathbf{U}_{\beta l}, {\bf{V}}_{\beta l}\}_{l=1}^L$.
Then there exists a nonzero $\mathbf{V}_{\alpha k}$ satisfying \eqref{eq:intercell_IA_beta} and \eqref{eq:intracell_IA_alpha} almost surely if 
\begin{equation*}
d_{\alpha k}\leq M_{\alpha}-\sum_{k'=1,k'\neq k}^K d_{\alpha k'}-\sum_{l'=1}^L d_{\beta l'}
\end{equation*}
since the total number of intra-cell and inter-cell interference dimensions is given by  $\sum_{k'=1,k'\neq k}^K d_{\alpha k'}+\sum_{l'=1}^L d_{\beta l'}$,
which yields the condition \eqref{eq:necessary(BS)} because of $M_{\alpha}\geq M_{\beta}$.

Now consider the case where $M_{\alpha}< M_{\beta}$. For this case, \eqref{eq:necessary(beta_intra)} becomes inactive from \eqref{eq:necessary(BS)}.
We construct $\{\mathbf{V}_{\alpha k}\}_{k=1}^K$ only satisfying \eqref{eq:intracell_IA_alpha}.
Then there exists a nonzero $\mathbf{V}_{\alpha k}$ satisfying \eqref{eq:intracell_IA_alpha} almost surely if 
\begin{equation*}
d_{\alpha k}\leq M_{\alpha}-\sum_{k'=1,k'\neq k}^K d_{\alpha k'},
\end{equation*}
which yields \eqref{eq:necessary(alpha_intra)}.
Then we construct $\{\mathbf{U}_{\beta l}\}_{l=1}^L$ satisfying \eqref{eq:intercell_IA_beta} and \eqref{eq:intracell_IA_beta} for given $\{{\bf{U}}_{\alpha k}, \mathbf{V}_{\alpha k}\}_{k=1}^K$ and $\{{\bf{V}}_{\beta l}\}_{l=1}^L$. %[KY]  $\{{\bf{U}}_{\alpha k}\}_{k=1}^K$ -> $\{{\bf{U}}_{\alpha k}, \mathbf{V}_{\alpha k}\}_{k=1}^K$
Then there exists a nonzero ${\bf{U}}_{\beta l}$ satisfying \eqref{eq:intercell_IA_beta} and \eqref{eq:intracell_IA_beta} almost surely if 
\begin{equation*}
d_{\beta l}\leq M_{\beta}-\sum_{l'=1,l'\neq l}^L d_{\beta l'}-\sum_{k'=1}^K d_{\alpha k'},
\end{equation*}
which yields the condition \eqref{eq:necessary(BS)} because of $M_{\alpha}< M_{\beta}$.

In conclusion, for both cases, $\{\mathbf{V}_{\alpha k}\}_{k=1}^K$ and $\{\mathbf{U}_{\beta l}\}_{l=1}^L$ satisfying \eqref{eq:intercell_IA_beta}  to \eqref{eq:intracell_IA_beta} exist almost surely if \eqref{eq:necessary(alpha_intra)} to \eqref{eq:necessary(BS)} are satisfied.

\subsubsection{Linear independence for the desired streams}
From the above construction, we can easily show that $\mathbf{U}_{\alpha k}$ and $\mathbf{V}_{\beta l}$ are not function of $\mathbf{H}_{\alpha k}$ and $\mathbf{H}_{\beta l}$ for all $k\in[1:K]$ and $l\in[1:L]$. 
Therefore, the conditions \eqref{eq:desired_alpha} and \eqref{eq:desired_beta} are satisfied almost surely.
In conclusion, Theorem \ref{the:sufficient_conditions} holds.

\subsection{Proof of Theorem \ref{the:divisible}}\label{subsec:proof_cor1}

In this subsection, we prove Theorem \ref{the:divisible}.
We construct $\{\mathbf{U}_{\alpha k}, \mathbf{V}_{\alpha k}\}_{k=1}^K$ and $\{\mathbf{U}_{\beta l}, \mathbf{V}_{\beta l}\}_{l=1}^L$ in the same manner used in the proof of Theorem \ref{the:sufficient_conditions}. %  

First, let us investigate the existence of $\{ {\bf{U}}_{\alpha k} \}_{k=1}^{K}$ and $\{ {\bf{V}}_{\beta l} \}_{l=1}^{L}$ satisfying \eqref{eq:intercell_IA_alpha}, which is rewritten as \eqref{eq:Reform_7a} from \eqref{eq:U_V_}.
Then denote the left-hand side of \eqref{eq:Reform_7a} by ${\mathbf{F}}_{k l}$ that is given by
\begin{align} \label{eq:ICImatrix}
{\mathbf{F}}_{k l}={\mathbf{G}}_{\alpha kl}^{(1)} + {\mathbf{G}}_{\alpha kl}^{(2)} {\mathbf{\overline V}}_{\beta l} + {\mathbf{\overline U}}_{\alpha k}^{\dag} {\mathbf{G}}_{\alpha kl}^{(3)}  + {\mathbf{\overline U}}_{\alpha k}^{\dag} {\mathbf{G}}_{\alpha kl}^{(4)} {\mathbf{\overline V}}_{\beta l}.
\end{align}
The Jacobian matrix of \eqref{eq:ICImatrix} with respect to variables $\{ \overline{\bf{U}}_{\alpha k} \}_{k=1}^{K}$ and $\{ \overline{\bf{V}}_{\beta l} \}_{l=1}^{L}$ is defined as
\begin{align} \label{eq:J}
{\bf{J}}  = \left[ {\frac{{\partial {\rm{vec}}({\bf{F}})}}{{\partial {\rm{vec}}(\overline {\bf{U}} ,\overline {\bf{V}} )}}} \right]
\end{align}
where ${\rm{vec}}({ \mathbf{F}})=[  {\rm{vec}}( { \mathbf{F}}_{11})^{T},{\cdots}, {\rm{vec}}({\mathbf{F}}_{KL})^{T} ]^T$ and 
\begin{align*} 
{\rm{vec}}({ \mathbf{\overline U}} , { \mathbf{\overline V}} ) 
=&[  {\rm{vec}}({ \mathbf{\overline U}}_{\alpha 1}^{\dag})^{T},{\cdots}, {\rm{vec}}({\mathbf{\overline U}}_{\alpha K}^{\dag})^{T}, \nonumber \\
 &{~~~~~~~~~~} {\rm{vec}}({ \mathbf{\overline V}}_{\beta 1})^{T},{\cdots},{\rm{vec}}({\mathbf{\overline V}}_{\beta L})^{T} ]^T.
\end{align*}

We then introduce the following lemma, which will be used for showing the existence of $\{ {\bf{U}}_{\alpha k} \}_{k=1}^{K}$ and $\{ {\bf{V}}_{\beta l} \}_{l=1}^{L}$ satisfying \eqref{eq:intercell_IA_alpha} almost surely.

\begin{lemma}[Existence of a specific channel realization]\label{lem:jacobian}
For the MIMO R-TDD cellular network, if \eqref{eq:cor1(divisible)} and \eqref{eq:cor1(variables_equations)} are satisfied, then there exist $\{{\mathbf{G}}^{(i)}_{\alpha kl}\}_{i\in[1:4],k\in[1:K],l\in [1:L]}$, denoted by $\{{\mathbf{G}}^{(i)*}_{\alpha kl}\}_{i\in[1:4],k\in[1:K],l\in [1:L]}$, such that $\mathbf{J}$ is non-singular for given $\{{\mathbf{G}}^{(i)*}_{\alpha kl}\}_{i\in[1:4],k\in[1:K],l\in [1:L]}$.
Furthermore, there exist $\{ {\bf{U}}_{\alpha k} \}_{k=1}^{K}$ and $\{ {\bf{V}}_{\beta l} \}_{l=1}^{L}$ satisfying \eqref{eq:intercell_IA_alpha} for given $\{{\mathbf{G}}^{(i)*}_{\alpha kl}\}_{i\in[1:4],k\in[1:K],l\in [1:L]}$.
\end{lemma}
\begin{IEEEproof}
Assume that the symmetric DoF $d_{\alpha}$ and $d_{\beta}$ satisfy \eqref{eq:cor1(divisible)} and \eqref{eq:cor1(variables_equations)}.
First we explain how to establish $\{{\mathbf{G}}^{(i)*}_{\alpha kl}\}_{i\in[1:4],k\in[1:K],l\in [1:L]}$ that guarantees non-singularity of $\mathbf{J}$.
Set ${\mathbf{G}}_{\alpha kl}^{(1)*}= {\mathbf{0}}$ and ${\mathbf{G}}_{\alpha kl}^{(4)*}= {\mathbf{0}}$ for all $k$ and $l$.
Then from \eqref{eq:cor1(divisible)} and \eqref{eq:ICImatrix}
\begin{align}\label{eq:ICIdivide}
&{{\bf{F}}_{kl}} = [ {\begin{array}{*{20}{c}}
{\overline {\bf{U}} _{\alpha k,1}^\dag }&{\overline {\bf{U}} _{\alpha k,2}^\dag }& \cdots &{\overline {\bf{U}} _{\alpha k,{A_k}}^\dag }
\end{array}} ]\left[ {\begin{array}{*{20}{c}}
{{\bf{G}}_{\alpha kl,1}^{(3)}}\\
{{\bf{G}}_{\alpha kl,2}^{(3)}}\\
 \vdots \\
{{\bf{G}}_{\alpha kl,{A_k}}^{(3)}}
\end{array}} \right] \nonumber \\
&{~~~~~~~~~~}+ \left[ {\begin{array}{*{20}{c}}
{{\bf{G}}_{\alpha kl,1}^{(2)}}\\
{{\bf{G}}_{\alpha kl,2}^{(2)}}\\
 \vdots \\
{{\bf{G}}_{\alpha kl,{B_l}}^{(2)}}
\end{array}} \right] [ {\begin{array}{*{20}{c}}
{{{\overline {\bf{V}} }_{\beta l,1}}}&{{{\overline {\bf{V}} }_{\beta l,2}}}& \cdots &{{{\overline {\bf{V}} }_{\beta l,{B_l}}}}
\end{array}} ], % [KY] ,?  
\end{align}
where ${\mathbf{\overline U}}_{\alpha k,i} \in \C ^{d_\beta \times d_\alpha}$, ${\mathbf{\overline V}}_{\beta l,j} \in \C ^{d_\alpha \times d_\beta}$, ${\mathbf{G}}_{\alpha kl,i}^{(3)} \in \C ^{d_\beta \times d_\beta} $, ${\mathbf{G}}_{\alpha kl,j}^{(2)} \in \C ^{d_\alpha \times d_\alpha}$ and $i \in [1:A_k]$, $j \in [1:B_l]$, $A_k = (N_{\alpha k} - d_{\alpha}) / {d_\beta}$, $B_l = (N_{\beta l}-d_{\beta})/d_{\alpha}$.

\begin{figure}
\begin{center}$
\centering\includegraphics[width=2.5in]{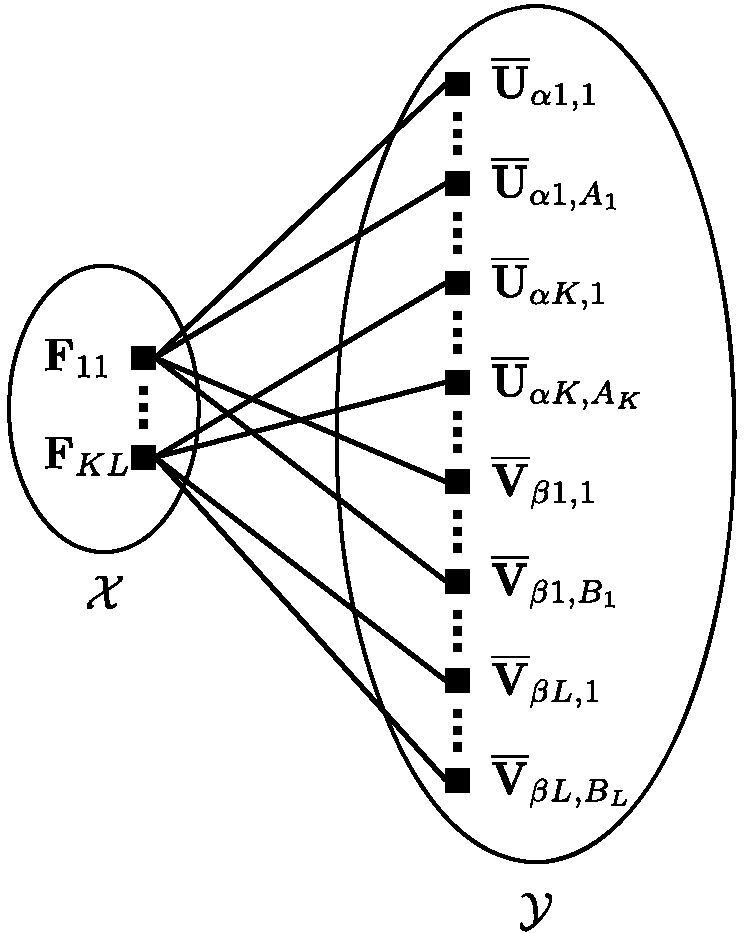}$
\end{center}
\caption{The bipartite graph of Lemma \ref{lem:jacobian}.}
\label{Fig:bipartite}
\end{figure}

We now introduce the bipartite graph $G(\mathcal{X},\mathcal{Y},\mathcal{E})$ depicted in Fig . \ref{Fig:bipartite}.
The vertex set $\mathcal{X}$ consists of $\mathbf{F}_{kl}$ for all $k\in[1:K]$, $l\in[1:L]$ and the vertex set $\mathcal{Y}$ consists of ${\mathbf{\overline U}}_{\alpha k,i}$, ${\mathbf{\overline V}}_{\beta l,j}$ for all $i \in [1:A_k]$, $j \in [1:B_l]$, $k\in[1:K]$, $l\in[1:L]$. % [KY] notation is changed X<->Y
There exist an edge between ${\mathbf{\overline U}}_{\alpha k,i}$ (or ${\mathbf{\overline V}}_{\beta l,j}$) and $\mathbf{F}_{kl}$ if ${\mathbf{\overline U}}_{\alpha k,i}$ (or ${\mathbf{\overline V}}_{\beta l,j}$) appears in the definition of $\mathbf{F}_{kl}$ in \eqref{eq:ICIdivide} where the edge set is denoted by $\mathcal{E}$. % [KY]+  notation of "E"
From Hall's marriage theorem\cite{book:graphtheory}, there is a complete matching in $G(\mathcal{X},\mathcal{Y},\mathcal{E})$ if and only if $\left| {\mathcal{X}}_{s} \right| \leq \left| N_{G}({\mathcal{X}}_{s}) \right|$ for  all possible subsets of vertices ${\mathcal{X}}_{s} \subseteq \mathcal{X}$, where $N_{G}({\mathcal{X}}_{s})$ denotes the set of neighbors of ${\mathcal{X}}_{s}$.
Notice that the above condition is satisfied by \eqref{eq:cor1(variables_equations)}, which guarantees the existence of a complete matching in $G(\mathcal{X},\mathcal{Y},\mathcal{E})$.

Suppose that $\mathcal{Y}_c\subseteq \mathcal{Y}$ denotes the set of vertices included in a complete matching. %[KY]  $\mathcal{X}_s$ -> $\mathcal{Y}_c$
Then set $\mathbf{G}^{(3)*}_{\alpha kl,i}=\mathbf{I}_{d_{\beta}}$ if $\overline {\bf{U}} _{\alpha k,i} $ is included in $\mathcal{Y}_c$, otherwise set   $\mathbf{G}^{(3)*}_{\alpha kl,i}=\mathbf{0}$ for all $i\in[1:A_k]$ and $k\in[1:K]$. %[KY]  $\mathcal{X}_s$ -> $\mathcal{Y}_c$
Similarly set $\mathbf{G}^{(2)*}_{\alpha kl,j}=\mathbf{I}_{d_{\alpha}}$ if $\overline {\bf{V}} _{\beta l,j} $ is included in $\mathcal{Y}_c$, otherwise set $\mathbf{G}^{(2)*}_{\alpha kl,j}=\mathbf{0}$ for all $j\in[1:B_l]$ and $l\in[1:L]$.  %[KY]  $\mathcal{X}_s$ -> $\mathcal{Y}_c$

As a result, $\mathbf{J}$ in \eqref{eq:J} becomes a block permutation matrix  for given $\{{\mathbf{G}}^{(i)*}_{\alpha kl}\}_{i\in[1:4],k\in[1:K],l\in [1:L]}$, which is non-singular.
Moreover, setting ${\mathbf{\overline U}}_{\alpha k}= \mathbf{0}$ for all $k\in[1:K]$ and ${\mathbf{\overline V}}_{\beta l}=\mathbf{0}$ for all $l\in[1:L]$ satisfies \eqref{eq:intercell_IA_alpha} for given $\{{\mathbf{G}}^{(i)*}_{\alpha kl}\}_{i\in[1:4],k\in[1:K],l\in [1:L]}$.
\end{IEEEproof}

By the same analysis in \cite{Bresler14, Razaviyayn12, Tingting13}, if there exists a specific set of $\{{\mathbf{G}}^{(i)*}_{\alpha kl}\}_{i\in[1:4],k\in[1:K],l\in [1:L]}$ that guarantees non-singular $\mathbf{J}$ and the existence of  $\{ {\bf{U}}_{\alpha k} \}_{k=1}^{K}$ and $\{ {\bf{V}}_{\beta l} \}_{l=1}^{L}$ satisfying \eqref{eq:intercell_IA_alpha}, then there exist $\{ {\bf{U}}_{\alpha k} \}_{k=1}^{K}$ and $\{ {\bf{V}}_{\beta l} \}_{l=1}^{L}$ satisfying \eqref{eq:intercell_IA_alpha} almost surely for a set of generic channel matrices $\{{\mathbf{G}}^{(i)}_{\alpha kl}\}_{i\in[1:4],k\in[1:K],l\in [1:L]}$.
From Lemma \ref{lem:jacobian}, we can find such  $\{{\mathbf{G}}^{(i)*}_{\alpha kl}\}_{i\in[1:4],k\in[1:K],l\in [1:L]}$ if \eqref{eq:cor1(divisible)} and \eqref{eq:cor1(variables_equations)} are satisfied.
In conclusion, if \eqref{eq:cor1(divisible)} and \eqref{eq:cor1(variables_equations)} are satisfied, $\{ {\bf{U}}_{\alpha k} \}_{k=1}^{K}$ and $\{ {\bf{V}}_{\beta l} \}_{l=1}^{L}$ satisfying \eqref{eq:intercell_IA_alpha} exist almost surely.

Proving the existence of  $\{\mathbf{U}_{\beta l}\}_{l=1}^L$ and $\{\mathbf{V}_{\alpha k}\}_{k=1}^K$ satisfying \eqref{eq:intercell_IA_beta} to \eqref{eq:intracell_IA_beta} is the same as that in Theorem \ref{the:sufficient_conditions}, which yields the conditions \eqref{eq:cor1(alpha_intra)} to \eqref{eq:cor1(BS)}.
Therefore, the conditions \eqref{eq:desired_alpha} and \eqref{eq:desired_beta} are satisfied almost surely from the same manner as Theorem \ref{the:sufficient_conditions}. % [KY]+ adding (7e) and (7f) condition
Therefore, Theorem \ref{the:divisible} holds.

%%%%% section: Procedure of making BM linear IA coder
\section{Iterative Construction of Precoding and Postcoding Matrices}\label{sec:IAbuilder}
Up to now, we considered necessary or sufficient conditions on the IA feasibility for MIMO R-TDD cellular networks. 
In this section, we focus on how to construct precoding and postcoding matrices assuming that $\left( {d_{\alpha 1}},\cdots ,{d_{\alpha K}}, {d_{\beta 1}},\cdots ,{d_{\beta L}} \right)$ satisfies the sufficient condition in Theorem \ref{the:sufficient_conditions}.
The overall procedure follows the same steps in Section \ref{subsec:proof_sufficient_conditions}.
Specifically, we apply the iterative method proposed in \cite{Gomadam11} for constructing $\{ {\bf{U}}_{\alpha k} \}_{k=1}^{K}$ and $\{ {\bf{V}}_{\beta l} \}_{l=1}^{L}$ satisfying \eqref{eq:intercell_IA_alpha}, i.e., Step 1.
Then we apply pseudo-inverse zero-forcing for constructing $\{\mathbf{V}_{\alpha k}\}_{k=1}^K$ and $\{\mathbf{U}_{\beta l}\}_{l=1}^L$ satisfying \eqref{eq:intercell_IA_beta} to \eqref{eq:intracell_IA_beta}, i.e., Step 2.

We assume that the transmit power of $P_{\alpha k}$, $k\in[1:K]$, is used for user $(\alpha, k)$ and the transmit power of $P_{\beta l}$, $l\in[1:L]$, is used for user $(\beta, l)$.
Hence, the transmit power of BS $\alpha$ is given by $P_{\alpha}=\sum_{i=1}^K P_{\alpha k}$.
We further assume equal power allocation between multiple streams for each user, i.e.,  $\frac{P_{\alpha k}}{d_{\alpha k}}$ is allocated for each of $d_{\alpha k}$ streams for user $(\alpha, k)$ and $\frac{P_{\beta l}}{d_{\beta l}}$ is allocated for each of $d_{\beta l}$ streams for user $(\beta, l)$.
In the following, we describe the proposed construction based on the normalized precoding and postcoding matrices. 
That is, the squared norm of each column vector of precoding and postcoding matrices is normalized by one.

\subsection{Construction of $\{{\bf{U}}_{\alpha k}\}_{k=1}^K$ and $\{{\bf{V}}_{\beta l}\}_{l=1}^L$} \label{subsec:construction1}
As mentioned above, we apply the iterative method in \cite{Gomadam11} for constructing $\{ {\bf{U}}_{\alpha k} \}_{k=1}^{K}$ and $\{ {\bf{V}}_{\beta l} \}_{l=1}^{L}$ satisfying \eqref{eq:intercell_IA_alpha}.
To describe the proposed construction, we define $\{{\bf{U}}_{\alpha k}^{[0]}\in\mathbb{C}^{N_{\alpha k} \times d_{\alpha k}}\}_{k=1}^K$, each of which is randomly chosen from the set of unitary matrices.
Let $\{ {\bf{U}}_{\alpha k}^{[i]} \}_{k=1}^{K}$ and $\{ {\bf{V}}_{\beta l}^{[i]} \}_{l=1}^{L}$ denote the precoding and postcoding matrices at the $i$th iteration respectively.
For the $i$th iteration, we first update ${\bf{V}}_{\beta l}^{[i]}$ from ${\bf{U}}_{\alpha k}^{[i-1]}$ and then sequentially update ${\bf{U}}_{\alpha k}^{[i]}$ from ${\bf{V}}_{\beta l}^{[i]}$. In the following, we state in details how to update ${\bf{V}}_{\beta l}^{[i]}$ and ${\bf{U}}_{\alpha k}^{[i]}$.

In order to update ${\bf{V}}_{\beta l}^{[i]}$ from ${\bf{U}}_{\alpha k}^{[i-1]}$, calculate the inter-cell interference covariance matrix for user ($\beta , l$) defined by 
\begin{equation}\label{eq:compute covariance matrix at TX}
{\bf{C}}_{\beta l}^{[i]}  = \sum\limits_{k = 1}^K {\frac{{P_{\alpha k} }}{{{d_{\alpha k}} }}{{\bf{G}}_{\alpha kl}^{\dag}} {\bf{U}}_{\alpha k}^{[i-1]} \left( {{{\bf{G}}_{\alpha kl}^{\dag}} {\bf{U}}_{\alpha k}^{[i-1]} } \right)^\dag  }.
\end{equation}
Then set 
\begin{equation}\label{eq:eigenvector of TX}
{\bf{V}}_{\beta l}^{[i]} \left[ {:,p} \right] =  \Gamma \left[ {{\bf{C}}_{\beta l}^{[i]} ,p} \right],\:\:\: \forall p \in [1 :d_{\beta l}]
\end{equation}
where $ \Gamma \left[ {{\bf{A}},n} \right]$ denotes the normalized eigenvector corresponding to the $n$th smallest eigenvalue of $\bf{A}$ and ${\mathbf{A}}[:,m]$ denotes the $m$th column vector of $\bf{A}$. 
Hence the resulting ${\bf{V}}_{\beta l}^{[i]}$ chooses the signal subspace that contains the least interference power.

Similarly, to update $ {\mathbf{U}}_{\alpha k}^{[i]} $ from ${\bf{V}}_{\beta l}^{[i]}$, calculate the inter-cell interference covariance matrix at user ($\alpha , k$) defined by
\begin{equation}\label{eq:compute covariance matrix at RX}
{\bf{C}}_{\alpha k}^{[i]}  = \sum\limits_{l = 1}^L {\frac{{P_{\beta l} }}{{{d_{\beta l}} }}{{\bf{G}}_{\alpha kl}} {\bf{V}}_{\beta l}^{[i]} \left( {{{\bf{G}}_{\alpha kl}} {\bf{V}}_{\beta l}^{[i]} } \right)^\dag}
\end{equation}
and set 
\begin{equation}\label{eq:eigenvector of RX}
{\bf{U}}_{\alpha k}^{[i]} \left[ {:,q} \right] = \Gamma \left[ {{\bf{C}}_{\alpha k}^{[i]},q} \right],\:\:\: \forall q \in [1:d_{\alpha k}]. 
\end{equation}
Table I summarizes the proposed iterative construction.

After the $\lambda$th iteration, the leakage interference power for user ($\alpha , k$) is given by
\[
I_{ \alpha k}^{[\lambda]} = {\rm{tr}} \bigg( {\mathbf{U}}_{\alpha k}^{[\lambda]\dag} {\mathbf{C}}_{\alpha k}^{[\lambda]} {\mathbf{U}}_{\alpha k}^{[\lambda]} \bigg)
\] 
where $\rm{tr}(\cdot)$ denotes the trace operation.
The total leakage interference power after the $\lambda$th iteration is then given by $I^{[\lambda]}=\sum_{k=1}^K I_{ \alpha k}^{[\lambda]}$.
It was shown in \cite{Gomadam11} that the total leakage interference power converges to zero as the number iterations increases, i.e., $I^{[\lambda]}\to 0$ as $\lambda\to \infty$. 
Hence we are able to decrease $I^{[\lambda]}$ as an arbitrarily small value for large enough $\lambda$, see also Fig. \ref{Fig:leakage}.

\begin{table} \label{table:alg}
\caption{Iterative construction for $\{ {\bf{U}}_{\alpha k} \}_{k=1}^{K}$ and $\{ {\bf{V}}_{\beta l} \}_{l=1}^{L}$.}
\hrulefill
\begin{algorithmic}[1]
\State {\textbf{Initialization}}: Fix $\lambda\in\mathbb{N}$ and set $\{ {\bf{U}}_{\alpha k}^{[0]} \}_{k=1}^{K}$, each of which is chosen from the set of unitary matrices uniformly at random.
\For{$i\in[1:\lambda]$}

	\State Calculate $\mathbf{C}_{\beta l}^{[i]}$ defined in \eqref{eq:compute covariance matrix at TX} and update ${\bf{V}}_{\beta l}^{[i]}$ as in \eqref{eq:eigenvector of TX}.
		
	\State Calculate $\mathbf{C}_{\alpha k}^{[i]}$ defined in \eqref{eq:compute covariance matrix at RX} and update ${\bf{U}}_{\alpha k}^{[i]}$ as in \eqref{eq:eigenvector of RX}.
\EndFor
\State{\textbf{Result}}: Set ${\bf{U}}_{\alpha k}={\bf{U}}^{[\lambda]}_{\alpha k}$ for all $k\in[1:K]$ and ${\bf{V}}_{\beta l}={\bf{V}}^{[\lambda]}_{\beta l}$ for all $l\in[1:L]$.
\end{algorithmic}
\hrulefill
\end{table}

\subsection{Construction of $\{\mathbf{V}_{\alpha k}\}_{k=1}^K$ and $\{\mathbf{U}_{\beta l}\}_{l=1}^L$} \label{subsec:construction2}

We now state in details how to construct $\{\mathbf{V}_{\alpha k}\}_{k=1}^K$ and $\{\mathbf{U}_{\beta l}\}_{l=1}^L$
for given $\{{\bf{U}}_{\alpha k}\}_{k=1}^{K}$ and $\{{\bf{V}}_{\beta l}\}_{l=1}^{L}$ satisfying \eqref{eq:intercell_IA_alpha}. 
That is, we assume that $\{{\bf{U}}_{\alpha k}\}_{k=1}^{K}$ and $\{{\bf{V}}_{\beta l}\}_{l=1}^{L}$ are constructed by the proposed iterative construction in Table I with large enough $\lambda$.

First, consider the case where $M_\alpha \geq M_\beta$. We build $\{{\bf{U}}_{\beta l}\}_{l=1}^{L}$ satisfying \eqref{eq:intracell_IA_beta}  as 
\[{[ {\begin{array}{*{20}{c}}
{{\bf{U}}_{\beta 1}^T}& \cdots &{{\bf{U}}_{\beta L}^T}
\end{array}} ]^T} = {({\bf{H}}_\beta ^{'\dag }{\bf{H}}_\beta ^{'})^{ - 1}}{\bf{H}}_\beta ^{'\dag }\]
where 
\[{\bf{ H}}_{\beta}^{'}  = [ {\begin{array}{*{20}c}
   {{\bf{H}}_{\beta 1} {\bf{V}}_{\beta 1}} & {{\bf{H}}_{\beta 2} {\bf{V}}_{\beta 2}} &  \cdots  & {{\bf{H}}_{\beta L} {\bf{V}}_{\beta L}}  \\
\end{array}} ].\]
After building $\{{\bf{U}}_{\beta l}\}_{l=1}^{L}$, we construct $\{{\bf{V}}_{\alpha k}\}_{k=1}^{K}$ satisfying \eqref{eq:intercell_IA_beta} and \eqref{eq:intracell_IA_alpha} such that 
${[ {\begin{array}{*{20}{c}}
{{\bf{V}}_{\alpha 1}}& \cdots &{{\bf{V}}_{\alpha K}}
\end{array}} ]} $
is set as the first $\sum_{k=1}^{K} d_{\alpha k}$ column vectors of ${\bf{H}}_{\alpha}^{'\dag}({\bf{H}}_{\alpha}^{' } {\bf{H}}_{\alpha}^{'\dag} )^{-1}$,
where 
\[{\bf{H}}_\alpha ^{'} = \left[ {\begin{array}{*{20}{c}}
{{\bf{U}}_{\alpha 1}^\dag {{\bf{H}}_{\alpha 1}}}\\
 \vdots \\
{{\bf{U}}_{\alpha K}^\dag {{\bf{H}}_{\alpha K}}}\\
{{{[ {\begin{array}{*{20}{c}}
{{\bf{U}}_{\beta 1}^{\dag T}}& \cdots &{{\bf{U}}_{\beta L}^{\dag T}}
\end{array}}]}^T}{{\bf{G}}_\beta }}
\end{array}} \right]\]
for given $\{{\bf{U}}_{\alpha k}\}_{k=1}^K$ and $\{{\bf{V}}_{\beta l},\mathbf{U}_{\beta l}\}_{l=1}^L$.

On the other hand, for $M_\alpha < M_\beta$, we first build $\{{\bf{V}}_{\alpha k}\}_{k=1}^{K}$ only satisfying \eqref{eq:intracell_IA_alpha} as
\[[ {\begin{array}{*{20}{c}}
{{{\bf{V}}_{\alpha 1}}}& \cdots &{{{\bf{V}}_{\alpha K}}}
\end{array}} ] ={\bf{H}}_{\alpha}^{''\dag}({\bf{H}}_{\alpha}^{''} {\bf{H}}_{\alpha}^{''\dag} )^{-1}\]
 where 
\[ {\bf{H}}_{\alpha}^{''}  = [ {\begin{array}{*{20}c}
   {{\bf{U}}_{\alpha 1}^\dag  {\bf{H}}_{\alpha 1} } & {{\bf{U}}_{\alpha 2}^\dag  {\bf{H}}_{\alpha 2} } &  \cdots  & {{\bf{U}}_{\alpha K}^\dag  {\bf{H}}_{\alpha K} }  \\
\end{array}} ].\]
Then construct $\{{\bf{U}}_{\beta l}\}_{l=1}^{L}$ satisfying \eqref{eq:intercell_IA_beta} and \eqref{eq:intracell_IA_beta} such that ${[ {\begin{array}{*{20}{c}}
{{\bf{U}}_{\beta 1}^T}& \cdots &{{\bf{U}}_{\beta L}^T}
\end{array}} ]^T} $ is set as the first $\sum_{l=1}^{L} d_{\beta l}$ row vectors of $({\bf{H}}_{\beta}^{'' \dag} {\bf{H}}_{\beta}^{''} )^{-1} {\bf{H}}_{\beta}^{''\dag}$, where 
\begin{align*}
&{{\bf{H}}_{\beta}^{''} } \nonumber \\
&= \bigg[ {\begin{array}{*{20}{c}}
{{{\bf{H}}_{\beta 1}} {{\bf{V}}_{\beta 1}}}& \cdots &{{{\bf{H}}_{\beta L}} {{\bf{V}}_{\beta L}}} & {{{\bf{G}}_\beta }\left[ {\begin{array}{*{20}{c}}
{{{\bf{V}}_{\alpha 1}}}& \cdots &{{{\bf{V}}_{\alpha K}}}
\end{array}} \right]}
\end{array}} \bigg].
\end{align*}

Finally, for both cases, we normalize the constructed $\{{\bf{V}}_{\alpha k}\}_{k=1}^{K}$ and $\{{\bf{U}}_{\beta l}\}_{l=1}^{L}$.
Specifically, we properly scale each column vector of $\mathbf{U}_{\beta l}$ (and $\mathbf{V}_{\alpha k}$) such that the squared norm of that column vector is equal to one. 
%As a result, the proposed construction of $\{\mathbf{V}_{\alpha k}\}_{k=1}^K$ and $\{\mathbf{U}_{\beta l}\}_{l=1}^L$ satisfy  \eqref{eq:intercell_IA_beta} to \eqref{eq:intracell_IA_beta}.

\subsection{Achievable Sum Rate} \label{subsec:sum_rate}
In this subsection, we derive the achievable sum rate by applying the precoding and postcoding matrices stated in Sections \ref{subsec:construction1} and \ref{subsec:construction2}.
Let the covariance matrix for the desired streams of user $(\alpha, k)$ as 
\begin{align*}
{\mathbf{C}}_{\alpha k}^{\sf{desire}} &= \frac{{{P_{\alpha k}}}}{{{d_{\alpha k}}}}{\bf{U}}_{\alpha k}^{\dag}{{\bf{H}}_{\alpha k}}{{\bf{V}}_{\alpha k}}{\left( {{\bf{U}}_{\alpha k}^{\dag}}{{\bf{H}}_{\alpha k}}{{\bf{V}}_{\alpha k}} \right)}^{\dag}.
\end{align*}
Also define
\begin{align*}
{\bf{C}}_{\alpha k}^{{\sf{inter}}} &= \sum\limits_{l = 1}^L {\frac{{{P_{\beta l}}}}{{{d_{\beta l}}}}{\bf{U}}_{\alpha k}^\dag {{\bf{G}}_{\alpha kl}}{{\bf{V}}_{\beta l}}{{\left( {{\bf{U}}_{\alpha k}^\dag {{\bf{G}}_{\alpha kl}}{{\bf{V}}_{\beta l}}} \right)}^\dag }}, \nonumber\\
{\bf {C}}_{\alpha k}^{{\sf{intra}}} &= \sum\limits_{i = 1,i \ne k}^K {\frac{{{P_{\alpha i}}}}{{{d_{\alpha i}}}}{\bf{U}}_{\alpha k}^\dag {{\bf{H}}_{\alpha k}}{{\bf{V}}_{\alpha i}}{{\left( {{\bf{U}}_{\alpha k}^\dag {{\bf{H}}_{\alpha k}}{{\bf{V}}_{\alpha i}}} \right)}^\dag }}, %[KY] , ?
\end{align*}
each of which are the inter-cell and intra-cell interference covariance matrices at user $(\alpha, k)$.
Then the achievable rate of user $(\alpha,k)$ is given by
\begin{align} \label{eq:r_alpha}
{R _{\alpha k}} = {\log _2}\det \left( {{{\bf{I}}_{{d_{\alpha k}}}} +  {{{\bf{C}}_{\alpha k}^{\sf{desire}}}} ({{{}{{\bf{I}}_{{d_{\alpha k}}}} + {\bf{C}}_{\alpha k}^{\sf{intra}} + {\bf{C}}_{\alpha k}^{\sf{inter}} }} )^{-1}} \right).
\end{align}

Similarly, the achievable rate of user $(\beta,l)$ is given by 
\begin{align} \label{eq:r_beta}
{R _{\beta l}} = {\log _2}\det \left( {{\bf{I}}_{{d_{\beta l}}}} +  {\bf{C}}_{\beta l}^{\sf{desire}} ({{}{{\bf{I}}_{{d_{\beta l}}}} + {\bf{C}}_{\beta l}^{\sf{intra}} + {\bf{C}}_{\beta l}^{\sf{inter}} } )^{-1} \right)
\end{align}
where
\begin{align*}
{\bf{C}}_{\beta l}^{\sf{desire}} &= \frac{{{P_{\beta l}}}}{{{d_{\beta l}}}}{\bf{U}}_{\beta l}^{\dag}{{\bf{H}}_{\alpha k}}{{\bf{V}}_{\beta l}}{\left( {{\bf{U}}_{\beta l}^{\dag}}{{\bf{H}}_{\beta l}}{{\bf{V}}_{\beta l}} \right)}^{\dag},\nonumber\\
{\mathbf{C}}_{\beta l}^{{\sf{intra}}} &= \sum\limits_{i = 1,i \ne l}^L {\frac{{{P_{\beta i}}}}{{{d_{\beta i}}}}{\bf{U}}_{\beta l}^\dag {{\bf{H}}_{\beta l}}{{\bf{V}}_{\beta i}}{{\left( {{\bf{U}}_{\beta l}^\dag {{\bf{H}}_{\beta l}}{{\bf{V}}_{\beta i}}} \right)}^\dag }},\nonumber\\
{\bf{C}}_{\beta l}^{{\sf{inter}}} &= \sum\limits_{k = 1}^K {\frac{{{P_{\alpha k}}}}{{{d_{\alpha k}}}}{\bf{U}}_{\beta l}^\dag {{\bf{G}}_\beta }{{\bf{V}}_{\alpha k}}{{\left( {{\bf{U}}_{\beta l}^\dag {{\bf{G}}_\beta }{{\bf{V}}_{\alpha k}}} \right)}^\dag }}.
\end{align*}
From \eqref{eq:r_alpha} and \eqref{eq:r_beta}, the achievable sum rate is given by
\begin{align} \label{eq:sum_rate}
R_{\sf{sum}} = \sum _{k=1}^{K} R_{\alpha k} + \sum _{l=1}^{L} R_{\beta l}. 
\end{align}

\begin{figure}
\begin{center}$
\centering\includegraphics[width=3.5in]{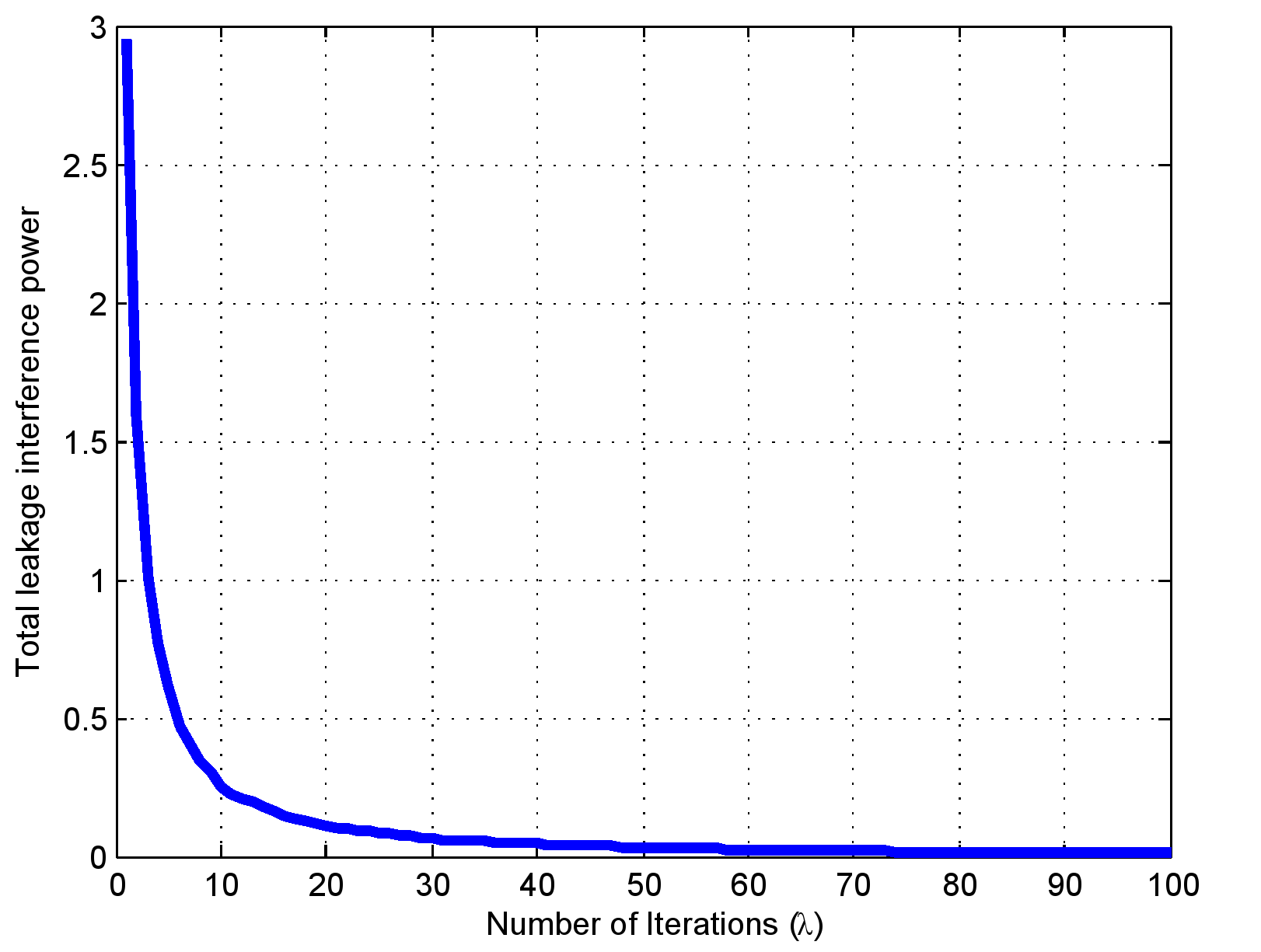}$
\end{center}
\caption{Total leakage interference power of the $( {12,\left( {8,8,8,8} \right) } ) \times ( {18,\left( {4,4,4} \right) } )$ MIMO R-TDD cellular network.}
\label{Fig:leakage}
\end{figure}
%da=2, db=3%
\begin{figure}
\begin{center}$
\centering\includegraphics[width=3.5in]{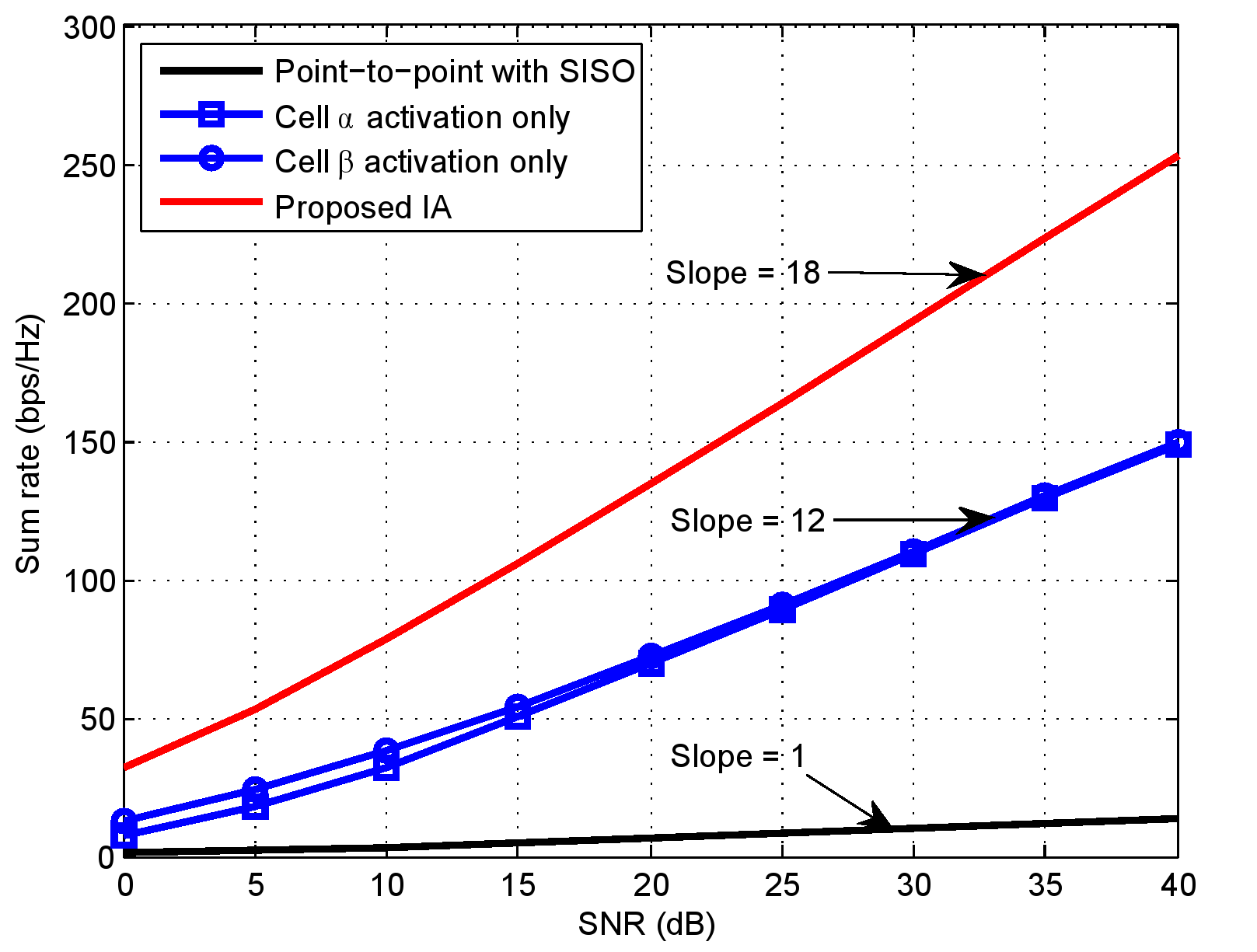}$
\end{center}
\caption{Sum rate of the $( {12,\left( {8,8,8,8} \right) } ) \times ( {18,\left( {4,4,4} \right) } )$ MIMO R-TDD cellular network with feasible sum DoF of $d_{\sf sum} = 18$.}
\label{Fig:sumrate}
\end{figure}

\subsection{Simulation}
In order to demonstrate the sum rate improvement achievable by the proposed construction at the finite SNR regime, we simulate the sum rate in Section \ref{subsec:sum_rate} for the ${(12,\left( {8,8,8,8} \right) } ) \times ( {18,\left( {4,4,4} \right) } )$ MIMO R-TDD cellular network. %[KY] signal-to-noise ratio (SNR) -> SNR
For this case, $d_{\alpha k}=3$ for all $k\in[1:4]$ and $d_{\beta l}=2$ for all $l\in[1:3]$ is feasible, which is used for simulation.
In simulation, we assume that channel coefficients are i.i.d. drawn from $\mathcal{CN}(0,1)$.
We then average out the performance over large enough channel realizations.

Figure \ref{Fig:leakage} plots the total leakage interference power $I^{[\lambda]}$ with respect to the number of iterations $\lambda$.
As shown in the figure, $I^{[\lambda]}$ quickly converges to zero as $\lambda$ increases, showing that the proposed iterative construction for $\{ {\bf{U}}_{\alpha k} \}_{k=1}^{K}$ and $\{ {\bf{V}}_{\beta l} \}_{l=1}^{L}$ works well for a moderate number of iterations. % [KY] a small or moderate -> a moderate

Figure \ref{Fig:sumrate} plots $R_{\sf{sum}}$ in \eqref{eq:sum_rate} with respect to SNR in dB scale. 
Specifically,  we set $P_{\alpha}=\mbox{SNR}$ and $P_{\beta l}=\mbox{SNR}$ for all $l\in[1:L]$. 
For comparison, we also plot the sum rate achievable by activating either cell $\alpha$ or cell $\beta$ only and the point-to-point capacity with single-antenna nodes.
We assume zero-forcing precoding and postcoding matrices for single-cell activation. 
For this system, optimal sum DoF is achieved by this proposed IA since Theorem \ref{the:necessary_conditions} implies $d_{\sf{sum}} \le 18$.
The result demonstrates that the proposed IA not only provides the optimal sum DoF but also significantly enhance the sum rate at the finite SNR regime. % [KY] the improved sum DoF -> the optimal sum DoF

\section{Concluding Remark}\label{sec:conclusions} %[KY]
In this paper, we established a necessary condition and a sufficient condition on IA feasibility for MIMO R-TDD cellular networks, which characterize the optimal sum DoF for many practical network configurations.
Our results demonstrate that interference can be effectively mitigated by the proposed one-shot linear IA using multiple antennas.
Although D-TDD or R-TDD systems have been actively studied, especially for heterogeneous cellular networks, there is still many remaining issues associated with IA in R-TDD systems. 
This work maybe extended to further arbitrary number of cells in R-TDD cellular networks.
For a practical deployment, an efficient IA scheme for heterogeneous R-TDD cellular networks needs to be studied further.

\ifCLASSOPTIONcaptionsoff
  \newpage
\fi

%\section*{Acknowledgements}

%\bibliographystyle{IEEEtran}
%\bibliography{IEEEabrv,IAreference}

\end{document}